\documentclass{aa}  

\usepackage{txfonts}
\usepackage[bookmarks=true,
            bookmarksopen=true,
            pdfstartview = FitV,
            pdfpagelayout = SinglePage,
            colorlinks = true,
            linkcolor=black,
            urlcolor = blue,
            pdfborder = {0 0 0}]{hyperref}

\begin{document}

   \title{The first simultaneous X-ray broad-band view of Mrk 110 with {\sl XMM-Newton} and {\sl NuSTAR}}

   \author{D.\ Porquet \inst{1}  
         \and
         J.~N.\ Reeves\inst{2,3}
          \and
          N.\ Grosso\inst{1}
           \and
          V.\ Braito\inst{3,2}
           \and            
           A.\ Lobban\inst{4}
         }
   \institute{Aix Marseille Univ, CNRS, CNES, LAM, Marseille, France 
              \email{delphine.porquet@lam.fr}
        \and  Department of Physics, Institute for Astrophysics and Computational
Sciences, The Catholic University of America, Washington, DC 20064, USA
          \and INAF, Osservatorio Astronomico di Brera, Via Bianchi 46 I-23807 Merate
 (LC), Italy
          \and European Space Agency (ESA), European Space Astronomy Centre (ESAC), E-28691 Villanueva de la Cañada, Madrid, Spain 
              }
   \date{Received , 2021; accepted , 2021}

 
  \abstract
  {Soft and hard X-ray excesses, compared to the continuum power-law shape between $\sim$2-10\,keV, are common features observed in the spectra of active galactic nuclei (AGN) and are associated with the accretion disc-corona system around the supermassive black hole. However, the dominant process at work is still highly debated and has been proposed to be either relativistic reflection or Comptonisation. 
  Such an investigation can be problematic for AGN having significant intrinsic absorption, either cold or warm, 
  which can severely distort the observed continuum. 
  Therefore, AGN with no (or very weak) intrinsic absorption along the line of sight, called bare AGN, 
  are the best targets to directly probe disc-corona systems. }
   {We aim to characterise the main X-ray spectral physical components from the bright bare Broad Line Seyfert 1 AGN Mrk\,110,
   and the physical process(es) at work in its disc-corona system viewed almost face-on.}
   {We perform the X-ray broad-band spectral analysis thanks to two simultaneous 
    {\sl XMM-Newton} and {\sl NuSTAR} observations performed on November 16-17 2019 and April 5-6 2020, we also use for the spectral analysis above 3 keV the deep {\sl NuSTAR} observation obtained in January 2017.}   
   {The broad-band X-ray spectra of Mrk\,110 are characterised by the presence of a prominent and absorption-free smooth soft X-ray excess, moderately broad \ion{O}{vii} and Fe\,K$\alpha$ emission lines and a lack of a strong Compton hump. 
  The continuum above $\sim$3\,keV is very similar at both epochs, while some variability (stronger when brighter) is present for the soft X-ray  excess. 
  A combination of soft and hard Comptonisation by a warm and hot corona, respectively, plus mildly relativistic disc reflection reproduce the broadband X-ray continuum very well. 
  The inferred warm corona temperature, kT$_{\rm warm}$\,$\sim$0.3\,keV, is similar to the values found in other sub-Eddington AGN, 
  whereas the hot corona temperature, kT$_{\rm hot}$\,$\sim$ 21--31\,keV (depending mainly on the assumed hot corona geometry), 
  is found to be in the lower range of the values measured in AGN. } 
{}
   \maketitle
%

\section{Introduction}\label{sec:Introduction}

In the standard picture, the emission of an active galactic nucleus
(AGN) stems from an accretion disc around a supermassive black hole (SMBH) with
mass spanning from a few millions to billions of solar masses.
X-ray spectra offer a unique potential to probe matter very close
to the black hole such as the disc-corona system. 
The X-ray spectra of AGN usually exhibit, in addition to a power-law continuum between $\sim$2 and 10\,keV, 
one or more of the following components : 
a soft X-ray excess below 2 keV, a Fe\,K${\alpha}$ line complex near 6.4 keV, and a Compton scattering hump 
peaking near 20--30\,keV. 
However, the geometry of the disc-corona system is still highly debated 
as well as the dominant emission process at work: 
either relativistic reflection resulting from the illumination of the accretion disc by a hot corona, 
or Comptonisation of seed photons from the accretion disc by a warm/hot corona 
\citep[e.g.,][]{Magdziarz98,P04a,Crummy06,Bianchi09,Fabian12,Done12,Gliozzi13,Petrucci18,Gliozzi20,Waddel20}. 
Moreover, some AGN exhibit warm absorbers along the line-of-sight that can  
severely complicate the X-ray data analysis. 
Therefore, bare AGN that show no (or very weak) X-ray
warm absorbers are the best targets to directly investigate the process(es) at work in the vicinity of SMBHs \citep{Porquet18}. \\

\begin{table*}[!t]
\caption{Observation log of the {\sl XMM-Newton} and {\sl NuSTAR} dataset analysed in this work. 
The (net) exposure times for XMM-Newton correspond to the pn ones.}
\begin{tabular}{c@{\hspace{15pt}}r@{\hspace{15pt}}l@{\hspace{15pt}}l@{\hspace{15pt}}c@{\hspace{15pt}}c@{\hspace{15pt}}c}
\hline \hline
Mission & Obs.\,ID & Obs.\,start (UTC) & Obs.\,end (UTC) & Exp. (ks) & count\,s$^{-1}$$^{(a)}$ \\
\hline 
{\sl NuSTAR}     & 60502022002 & 2019 November 16 -- 03:31:09 & 2019 November 18 -- 00:56:09 & 86.8 & 0.76 \\
{\sl XMM-Newton} &  0852590101 & 2019 November 17 -- 09:02:57 &  2019 November 17 -- 21:24:37 & 44.5 & 21.2\\
\hline
{\sl NuSTAR}     & 60502022004 & 2020 April 5 -- 14:26:09 & 2020 April 7 -- 13:26:09& 88.7 &  0.66 \\
{\sl XMM-Newton} & 0852590201 & 2020 April 6 -- 22:26:50 & 2020 April 7 -- 11:55:10 & 48.5 & 16.5 \\
\hline
{\sl NuSTAR}     &  60201025002 & 2017 January 23 -- 19:11:09 & 2017 January 28 06:41:09 & 184.6 &  1.08 \\
\hline
\end{tabular}
\label{tab:log}
\flushleft
\small{\textit{Notes.} $^a$Source count rate over 0.3--10 keV for {\sl XMM-Newton}/pn and over 3--79\,keV for {\sl NuSTAR}.} 
\end{table*}

The bright bare AGN \object{Mrk\,110} (z=0.035291) has been classified as a Narrow Line Seyfert 1 (NLS1) due to its 
 relatively narrow optical emission lines ($<$2000 km/s FWHM) emitted by the broad-line region (BLR) 
 \citep{Osterbrock85,Goodrich89}. 
 However, two much broader and redshifted line components are also observed ($\sim$5000--6000\,km\,s$^{-1}$ and $\sim$3340\,km\,s$^{-1}$; \citealt{Bischoff99,Veron-cetty07}). 
In addition, contrary to NLS1s, Mrk 110 displays an unusually large 
[\ion{O}{iii}]$\lambda$5007/H$\beta$ ratio and very weak \ion{Fe}{ii} 
emission, which is more consistent with a Broad Line Seyfert\,1 (BLS1),
and indicates a low Eddington ratio \citep{Boroson92,Grupe04,Veron-cetty07}. 
 Its X-ray photon index, soft X-ray excess strength and X-ray variability are similar to that found 
for BLS1s \citep{P04a,Boller07,Piconcelli05,Zhou10b,Ponti12,Gliozzi20}. 
It hosts a SMBH with a mean mass value of M$_{\rm grav}$=1.4$\pm$0.3$\times$10$^{8}$\,M$_{\odot}$ 
measured from the detection of gravitational redshifted emission in the variable component of all of the broad optical lines \citep{Kollatschny03,Liu17}. 
In contrast to the virial method, for which a much lower black hole mass has been estimated, 
this gravitational method is independent of any assumption on the BLR geometry, such as its inclination angle. 
 Indeed, as argued by \cite{Liu17}, the large discrepancy between M$_{\rm vir}$ and M$_{\rm grav}$ for Mrk\,110 is explained by the 
use of a virial factor of $f_{\rm FWHM}$\,$\sim$1 to infer M$_{\rm vir}$, which is equivalent for a geometrically thin disc-like BLR to an inclination angle $\theta$=30 degrees ($f_{\rm FWHM}$=1/(4\,sin$^2\theta$); the squared of Eq.\,4 of \citealt{Decarli08}).
Therefore, the $\sim$8 times lower virial mass of Mrk\,110 compared to its gravitational mass is explained by $f_{\rm FWHM}$\,$\approx$8--16 \citep{Liu17}, that is to say a BLR-disc system seen almost face-on with an inclination of 7--10 degrees (see also \citealt{Bian02,Kollatschny03}). 
 Consistent black hole mass values to M$_{\rm grav}$ have been inferred for Mrk\,110 
from different independent methods, such as optical spectro-polarimetric observations 
\citep{Afanasiev19}, X-ray excess variance \citep{Ponti12}, and X-ray scaling method \citep{Gliozzi11,Williams18}.
 From its optical/X-ray properties, black hole mass and accretion rate, Mrk\,110 is then a BLS1.\\

\cite{Reeves21b} found that the RGS spectra of Mrk\,110 obtained between 2004 and 2020  
exhibit no significant intrinsic X-ray warm absorption with an upper limit for its column density of only  2.6$\times$10$^{20}$\,cm$^{-2}$. 
This points out that Mrk\,110 is a bare AGN whatever its flux level is. 
Moreover, they confirm the presence of a broad \ion{O}{vii} soft X-ray emission line, first identified by \cite{Boller07}. 
For the first time, \cite{Reeves21b} report that the  \ion{O}{vii} line flux varies significantly with the soft X-ray continuum flux level, 
being brightest when the continuum flux is highest, similar to the reported behaviour of the optical \ion{He}{ii} line \citep{Veron-cetty07}. 
This \ion{O}{vii} line originates from the accretion disc at a distance of a few tens of gravitational radii.   
From the spectral analysis a very stringent constraint on the inclination angle of the accretion disc
has been measured, 9.9$^{+1.0}_{-1.4}$\,degrees, which is consistent with an almost face-on view, as also inferred from the optical emission lines emitted by the disc-like BLR \citep{Bian02,Kollatschny03,Liu17}. \\

In this paper, we report the X-ray broad-band spectral analysis of the first 
two simultaneous {\sl XMM-Newton} and {\sl NuSTAR} observations of Mrk\,110, performed on November 16-17 2019
and April 5-6 2020. 
In section~\ref{sec:obs}, the data reduction and analysis method of the dataset are presented. 
The data analysis above 3\,keV also uses  the deep {\sl NuSTAR} observation performed in January 2017 (section~\ref{sec:above3keV}). 
The X-ray broad-band spectra analysis using a combination of Comptonisation and relativistic reflection models is reported (section~\ref{sec:Compt/rel}).  
Finally, the main results are summarised and discussed in section~\ref{sec:discussion}.

\section{Observations, data reduction and analysis method}\label{sec:obs}

\subsection{{\sl XMM-Newton} and {\sl NuSTAR} data reduction}

The log of the {\sl XMM-Newton} and {\sl NuSTAR} simultaneous observations of Mrk\,110 
(NuSTAR cycle-5; PI: D.\ Porquet) used in this work is reported in Table~\ref{tab:log}. 
For the data analysis above 3\,keV ($\S$\ref{sec:above3keV}), 
we also use the deep {\sl NuSTAR} observation performed in January 2017.

The {\sl XMM-Newton}/EPIC event files were
reprocessed with the Science Analysis System (SAS, version 18.0.0), applying the
latest calibration available on 2021 February 24. Due to the high source
brightness, the EPIC instruments were operated in the Small Window mode. 
Only the EPIC/pn \citep{Struder01} 
data are used (selecting the event patterns 0--4, that is to say,
single and double events) since they do not suffer from pile-up (contrary to the MOS data)
and they have a much better sensitivity above $\sim$6\,keV. 
The pn spectra were extracted from a circular region centered on Mrk\,110, with a radius of
35${\arcsec}$ to avoid the edge of the chip. The background
spectra were extracted from a rectangular region in the lower part of
the small window that contains no (or negligible) source photons.  
 The total net exposure times, obtained after the correction for dead time and background flaring, are reported
 in Table~\ref{tab:log}. 
Redistribution matrix files (rmf) and ancillary response files (arf) were generated with the SAS tasks
{\sc rmfgen} and {\sc arfgen}, and were binned in order to over-sample the instrumental resolution by at
least a factor of three, with no impact on the fit results.  
As shown in Lobban et al.\ (2021, in prep.), there is some flux
variability during each {\sl XMM-Newton} observation and between them. 
However, the spectral variability within any single observation is slow and moderate,
therefore we use the pn time-averaged spectra for each epoch. 
 Finally, the background-corrected pn spectra were binned in order to have a signal-to-noise 
 ratio greater than four in each spectral channel. 

{\sl NuSTAR} \citep{Harrison13} observed Mrk\,110 with its two
co-aligned X-ray telescopes with corresponding Focal Plane Modules A 
(FPMA) and B (FPMB).  The level 1 data products were
processed with the {\sl NuSTAR} Data Analysis Software (NuSTARDAS)
package (version 2.0.0; 2020 June 19). Cleaned event files (level 2 data products) were
produced and calibrated using standard filtering criteria with the
\textsc{nupipeline} task and the calibration files available in the
{\sl NuSTAR} calibration database (CALDB: 20210210).
Extraction radii for both the source and the background spectra were $75$
arcseconds. The corresponding net exposure time for the observations with FPMA 
are reported in Table~\ref{tab:log}. 
 The processed rmf and arf files are provided on a linear grid of 40 eV steps. 
As the FWHM energy resolution of {\sl NuSTAR} is 400\,eV below $\sim$50 keV and increases to 1\,keV at 86\,keV \citep{Harrison13}, 
we re-binned the rmf and arf files in energy and channel space by a factor of 4 to over-sample the instrumental energy resolution
by at least a factor of 2.5.  
The background-corrected {\sl NuSTAR} spectra were finally binned in order to have a signal-to-noise ratio greater than five in each
spectral channel.

\subsection{Spectral analysis method}\label{sec:method}

The {\sc xspec} v12.11.1 software package \citep{Arnaud96}  
was used for the spectral analysis.
Since \cite{Reeves21b} found that there is no additional absorption compared to the Galactic one, 
we fix it in all forthcoming modelling at 1.27$\times$10$^{20}$\,cm$^{-2}$ \citep{HI4PI}.
We used the X-ray absorption model {\sc tbnew (version 2.3.2)} from
\cite{Wilms00}, applying their interstellar medium (ISM) elemental abundances and
the cross-sections from \cite{Verner96}. 
We allow for cross-calibration uncertainties between the two {\sl NuSTAR}
spectra and the {\sl XMM-Newton}/pn spectra by including in
the fit a cross-normalization factor  
for the pair of {\sl NuSTAR} FPMA and FPMB spectra, with respect to the pn spectra.  
Except in $\S$\ref{sec:main}, the inclination angle
of the accretion disc is fixed at 9.9\,degrees \citep{Reeves21b}. 
We use $\chi^{2}$ minimization throughout, quoting errors with 90\% confidence intervals 
for one interesting parameter ($\Delta\chi^{2}$=2.71). Default values of H$_{\rm 0}$=67.66\,km\,s$^{-1}$\,Mpc$^{-1}$,
$\Omega_{\rm m}$=0.3111, and $\Omega_{\Lambda}=0.6889$ are assumed
\citep{Planck20}.

\begin{figure}[t!]
\begin{tabular}{cc}
\includegraphics[width=0.9\columnwidth,angle=0]{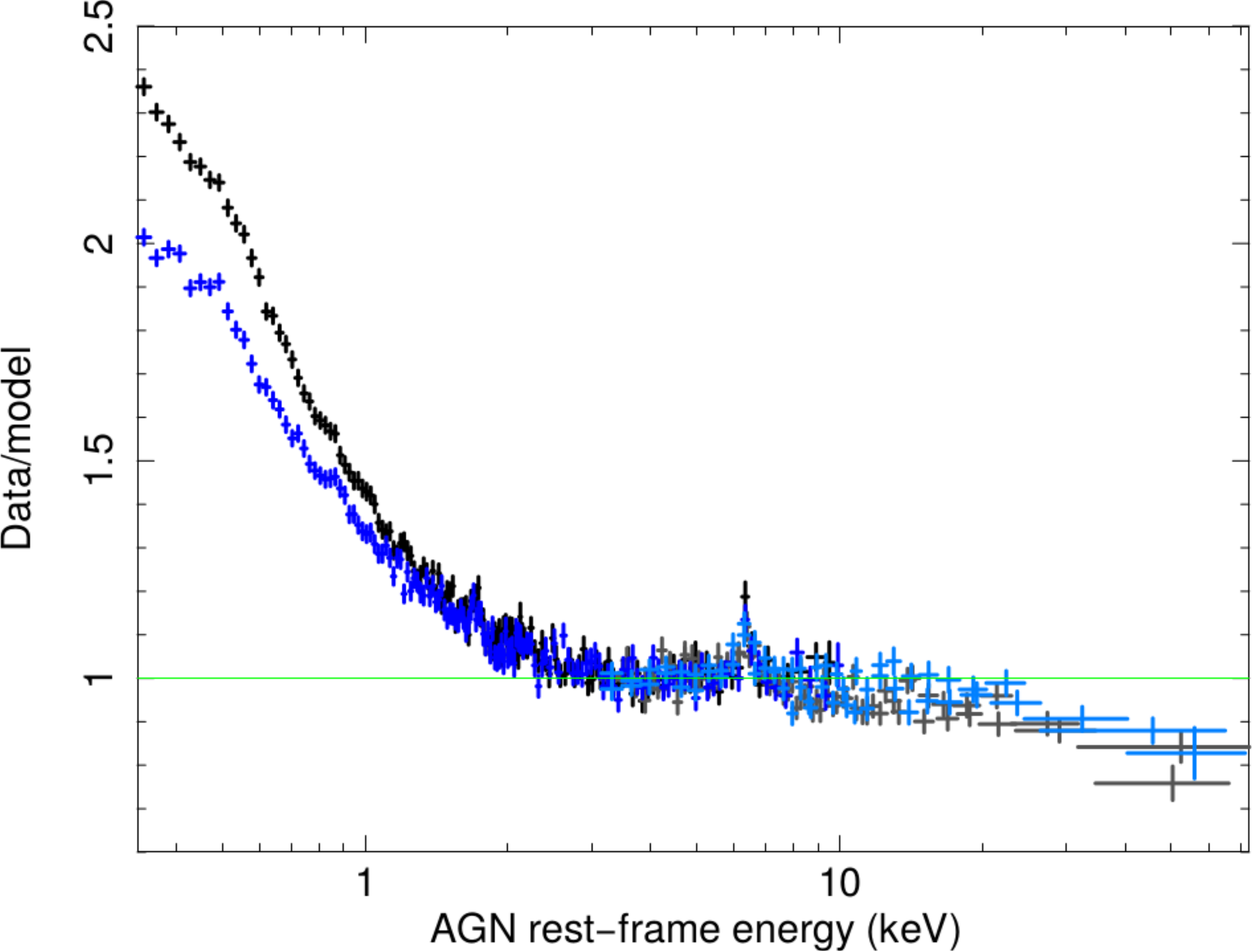}\\
\includegraphics[width=0.9\columnwidth,angle=0]{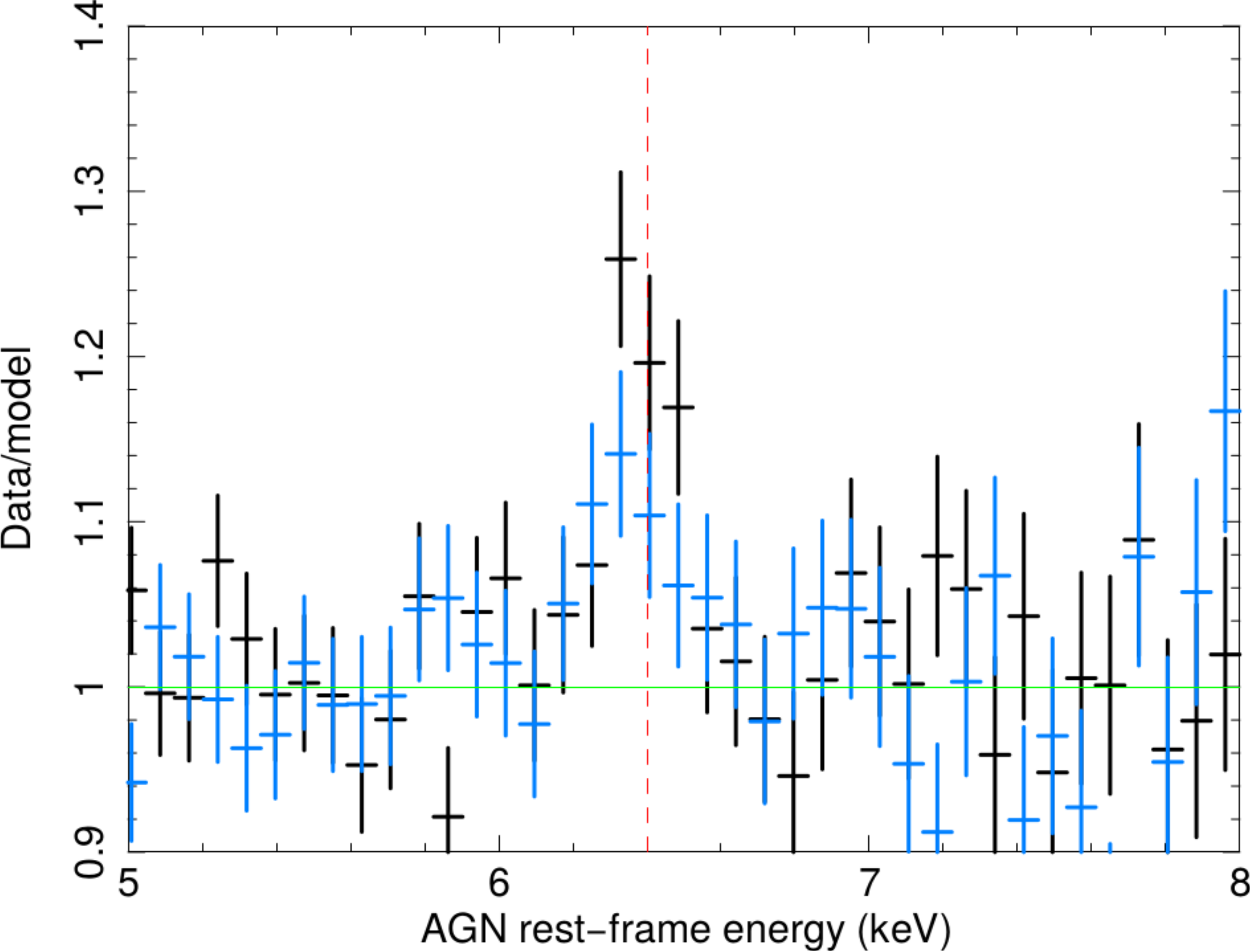}
\end{tabular}
	\caption{Data/model ratio of the two simultaneous {\sl XMM-Newton}/pn and {\sl NuSTAR} spectra of Mrk\,110 fit  with a  power-law (with Galactic absorption) model over the
 3--5\,keV energy range. 
 Black: 2019 {\sl XMM-Newton}/pn, dark grey: 2019 {\sl NuSTAR}, 
 blue: 2020 {\sl XMM-Newton}/pn, light blue: 2020 {\sl NuSTAR}. 
{\it Top panel}: Extrapolation over the full broad-band X-ray energy range. 
{\it Bottom panel}: Zoom on the Fe\,K$\alpha$ line. Only the pn spectra (using here a 75\,eV spectral binning)  
are reported for purposes of clarity. 
The vertical dashed red line corresponds to 6.4 keV.}  
\label{fig:3-5keV}
\end{figure}

\section{Main X-ray spectral components of the 2019 and 2020 XMM-Newton and NuSTAR observations}\label{sec:main}

In order to characterise the main X-ray components of the
spectra, we fit the two simultaneous {\sl XMM-Newton}/pn and {\sl NuSTAR} spectra 
between 3--5\,keV (AGN rest-frame) using a power-law with Galactic absorption. 
We find photon indices of 1.70$\pm$0.07 and 1.69$\pm$0.07 for November 2019 and April 2010, respectively.
As illustrated in Fig.~\ref{fig:3-5keV} (top panel), the extrapolation of this fit over the full energy range shows a prominent and absorption-free smooth soft X-ray excess below 2\,keV which is variable between the two 
epochs, with the highest flux observation (2019) exhibiting the largest soft X-ray excess. 
Mrk\,110 does not display any Compton hump, which suggests a weak (or even a lack of) relativistic reflection component; 
 Mrk\,110 has a flat hard X-ray shape with a notable deficit above $\sim$30\,keV that indicates a low high-energy cut-off, 
 that is to say, a low hot corona temperature.\\

The Fe\,K$\alpha$ line profile is found to be consistent between these two epochs, 
though the flux of the latter observation is lower (Fig.~\ref{fig:3-5keV}, bottom panel). 
Aiming to characterise this Fe\,K$\alpha$ complex, we first fit the two pn spectra between 3 and 10\,keV 
with a power-law with Galactic absorption and a Gaussian line. 
The energy and width ($\sigma$) of the line are tied between both observations.
We find a good fit with E=6.36$\pm$0.03\,keV and $\sigma$\,$\leq$0.106\,keV  ($\chi^{2}$/d.o.f.=585.1/669).
The equivalent widths (EW) are 58$\pm$16\,eV and 41$^{+16}_{-15}$\,eV for 2019 and 2020, respectively.
The mean value of the line energy is slightly lower than that expected from the neutral Fe\,K$\alpha$ 
fluorescence line in the AGN rest-frame.
We checked that this is not due to a gain issue by analysing other AGN observed at similar periods using the small window mode.
Even if the line width is not constrained, its high upper limit and the slightly lower line energy might indicate
that a mild relativistic reflection component is present. The upper limit to the line width corresponds to 
a FWHM velocity width, of $<$12000\,km\,s$^{-1}$, which is compatible with the widths of the accretion disc \ion{O}{vii} emission  line in the high resolution RGS spectra, for which the inner radius of the line emission is inferred to lie between about 20--100 $R_{\rm g}$ from the black hole \citep{Reeves21b}. We note that since the accretion disc is seen almost face-on, the width of lines emitted above a few 10s of $R_{\rm g}$ appears much narrower than for systems viewed at higher inclination angles, such as found in the bare AGN \object{Ark\,120} for the which the system is observed with an inclination angle of $\sim$30\,degrees  \citep{Porquet18,Porquet19}. This explains the moderate value of the width upper limit for Mrk\,110. 
Moreover, the ionisation parameter of the medium 
from which the \ion{O}{vii} emission line originates, log\,$\xi$\,$\sim$1.2, is expected to also give rise to a neutral Fe\,K emission line. 

Therefore, we now fit the two pn spectra between 3 and 10\,keV using a power-law with Galactic absorption, 
a narrow neutral Fe\,K$\alpha$ core ($E_{\rm n}$=6.4\,keV and $\sigma_{\rm n}$=0\,eV), 
 and a disc line model ({\sc relline}; \citealt{Dauser10,Dauser13}) to account for the mildly broad line. 
 For the latter component, the energy was fixed at 6.4\,keV and the spin at zero 
(since it is not constrained).
All parameters of the Fe\,K$\alpha$ narrow component and the emissivity index 
for the mild relativistic Fe\,K line component 
are tied between the two epochs. 
From this modelling ($\chi^{2}$/d.o.f.=585.4/668), 
we infer an accretion disc emissivity index lower than 2.1 and a disc inclination lower than 24.6 degrees, which is
consistent with the inclination angle inferred from the variable \ion{O}{vii} soft X-ray emission line ($\theta=9.9^{+1.0}_{-1.4}$) arising from the accretion disc too \citep{Reeves21b}. 
For the narrow core of the Fe\,K$\alpha$ line, we measure a EW$_{\rm n}$\,$\lesssim$ 20\,eV, 
which is in the lower range of the values found for  
 type-1 AGN ($\sim$30--200\,eV; \citealt{Liu10,Shu10,Fukazawa11,Ricci14}). 
This is consistent with the very weak covering factor of 0.06 for the putative torus, which is inferred from the ratio of the infrared to the bolometric luminosity of the source \citep{Ezhikode17}. Moreover, the torus covering factor measured for Mrk\,110 is one of the lowest reported in \cite{Ezhikode17}'s sample; this is also in agreement with the lack of Compton hump. 
The equivalent widths of the moderately broad Fe\,K$\alpha$ line are 
$EW_{\rm b}$=61$^{+23}_{-27}$\,eV and $EW_{\rm b}$=43$^{+22}_{-27}$\,eV for 2019 and 2020, respectively. 
Likewise, if we allow the inner disc radius to be free to vary and assume a standard value of three 
for the disc emissivity index, as well as fixing the inclination angle at 9.9 degrees ($\chi^{2}$/d.o.f.=585.1/668), we infer  
 $R_{\rm in}$=120$^{+263}_{-67}$\,R$_{\rm g}$. 
This strengthens that the moderately broad Fe line could originate from the accretion disc but not too close to ISCO, as was found for the \ion{O}{vii} soft X-ray emission line \citep{Reeves21b}. 
Its origin from the outer accretion disc is also consistent with the timing analysis reported in Lobban et al. (2021, in prep.)
where a hint of an extra variability is found at the Fe\,K$\alpha$ complex energy range in the fractional variability spectra.

\begin{figure}[t!]
\begin{tabular}{c}
 \includegraphics[width=0.9\columnwidth,angle=0]{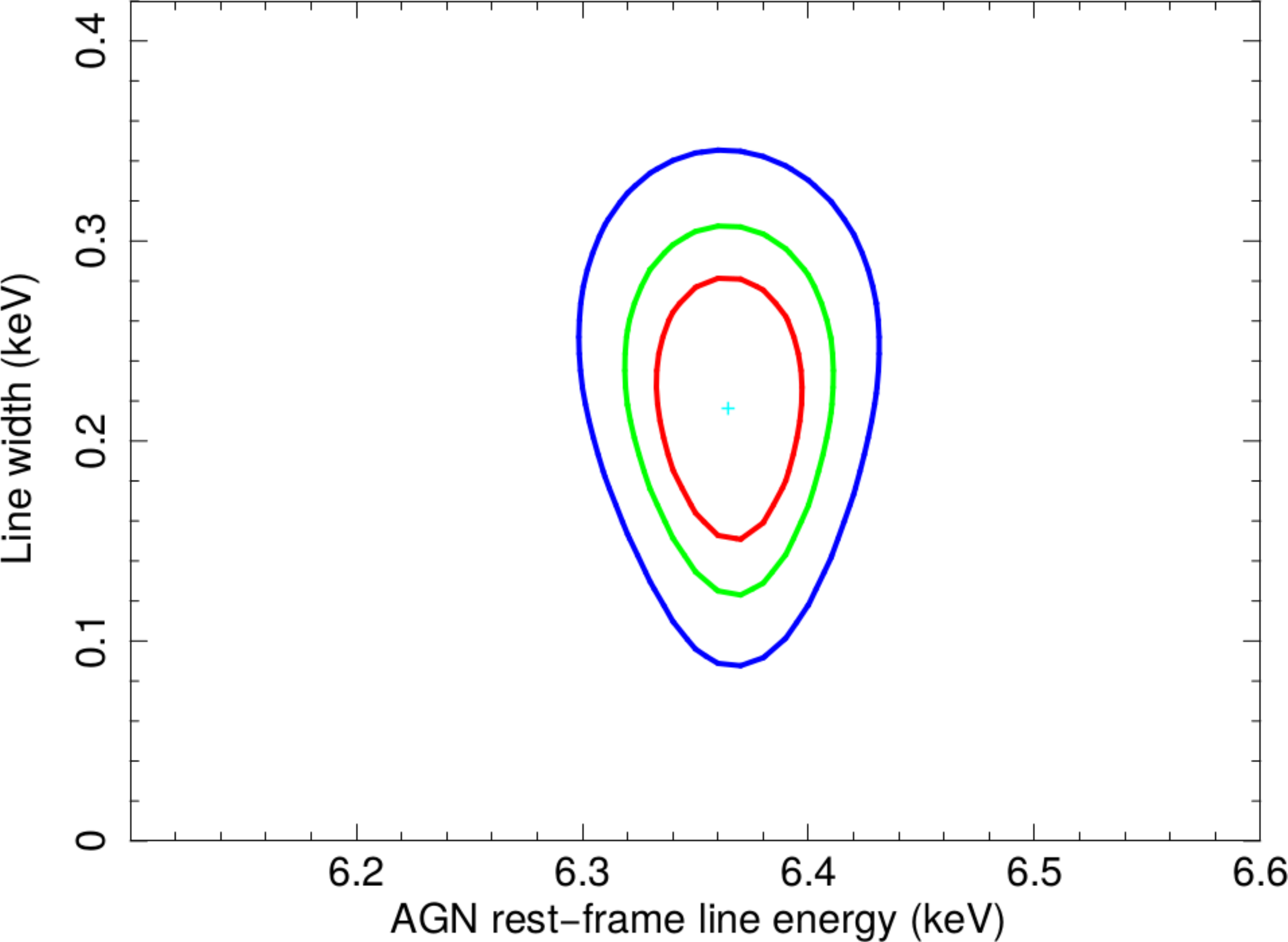} \\
\end{tabular}
\caption{Contour plot of the Fe\,K$\alpha$ Gaussian line width (keV) versus its energy (keV), assuming an underlying cut-off power-law continuum over the 3--79\,keV energy range (see $\S$\ref{sec:above3keV}). The red, green and blues curves show the confidence levels at 68\% ($\Delta\chi^2$=2.3), 90\% ($\Delta\chi^2$=4.61) and 99\% ($\Delta\chi^2$=9.21), respectively.} 
\label{fig:2D}
\end{figure}

\begin{figure}[t!]
\begin{tabular}{c}
 \includegraphics[width=0.9\columnwidth,angle=0]{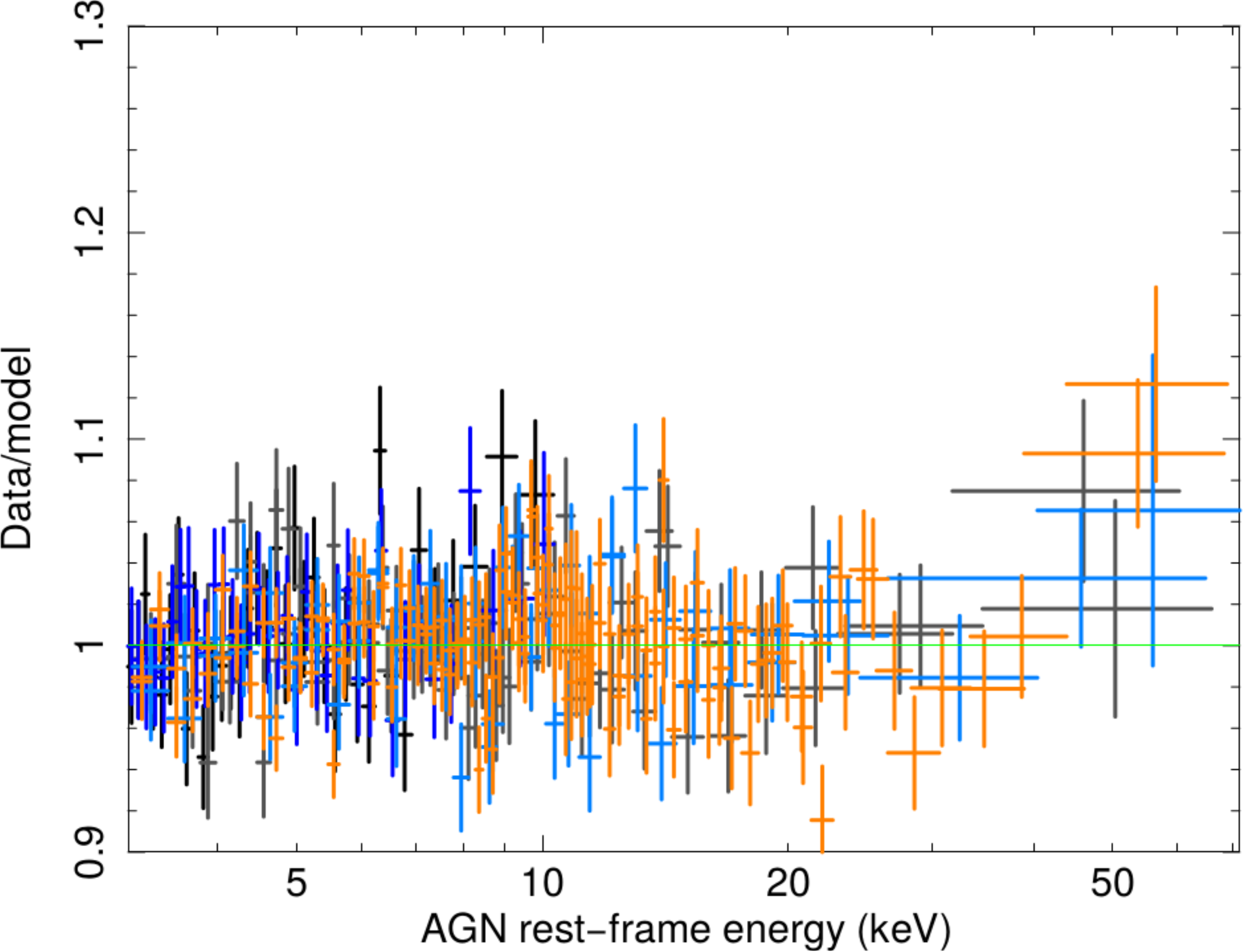} \\
\end{tabular}
\caption{Data-to-model ratio of the fit above 3\,keV with the {\sc reflkerr} relativistic reflection model, using a primary Comptonisation continuum shape and assuming a slab geometry for the hot corona. 
Black: 2019 {\sl XMM-Newton}/pn, dark grey: 2019 {\sl NuSTAR}, blue: 2020 {\sl XMM-Newton}/pn, light blue: 2020 {\sl NuSTAR} and orange: 2017 {\sl NuSTAR}.} 
\label{fig:reflabove3keV}
\end{figure}

\section{Spectral analysis above 3\,keV at three epochs: 2017, 2019 and 2020}\label{sec:above3keV}

In this section, we focus on the hard X-ray shape of Mrk\,110 
using the simultaneous {\sl XMM-Newton} and {NuSTAR} data above 3\,keV 
 to prevent the fit to be driven by the soft X-ray emission. 
 We also here include the deep {\sl NuSTAR} observation of 
Mrk\,110 performed in January 2017 \citep{Ezhikode20,Panagiotou20}.\\

 We first fit the data of these three epochs with a phenomenological model 
combining a power law ($\Gamma$) continuum with an exponential cut-off 
at high energy ($E_{\rm cut}$) and a Gaussian line. 
The line energy and its width are assumed to be constant between the three epochs, 
while its normalisation is allowed to vary. 
A very good fit is found as reported in Table~\ref{tab:above3keV}. 
Thanks to the use of both {\sl XMM-Newton} and {\sl NuSTAR} data over the 3--79\,keV and 
the deep 2017 {\sl NuSTAR} data, both the energy and width of the Fe\,K$\alpha$ 
line are well constrained, as illustrated by the 2D contour plot displayed in Fig.~\ref{fig:2D}. 
The AGN rest-frame line energy is slightly redshifted toward 6.4\,keV as previously found  
with the pn data, and line width corresponds to a FWHM of 22\,000$^{+7800}_{-4400}$\,km\,s$^{-1}$. 
This confirms that a mild relativistic reflection component is present. \\

 Now, we consider more physical models in order to determine both the hot corona properties 
and the disc reflection contribution.
 For this purpose, we use models that combine the primary continuum shape (hot corona)
and the relativistic reflection contribution: {\sc relxill/relxillcp}\footnote{\href{https://www.sternwarte.uni-erlangen.de/~dauser/research/relxill/}{https://www.sternwarte.uni-erlangen.de/$\sim$dauser/research/relxill/}} \citep[version 1.4.3;][]{Dauser10,Garcia16}
and {\sc reflkerr}\footnote{The usage notes as well as the full description of the models 
and their associated parameters are reported at \href{https://users.camk.edu.pl/mitsza/reflkerr/reflkerr.pdf}{https://users.camk.edu.pl/mitsza/reflkerr/reflkerr.pdf} model.} \citep[version 2019;][]{Niedzwiecki19}.

Since the lamppost geometry does not allow us to obtain satisfactory fits, 
we only report below the results for a coronal geometry.  
The single power-law disc emissivity index ($q$; with emissivity $\propto R^{-q}$) is fixed at the standard value of three, 
while the inner radius of the reflection component ($R_{\rm in}$, expressed in units of $R_{\rm g}$) is allowed to vary. 
The inclination of the accretion disc is fixed at 9.9 degrees and the iron abundance is set to unity.  
Since the spin value is not constrained, we fix it at zero.

\begin{table}[t!]
  \caption{Simultaneous fits above 3\,keV of the 2019 and 2020 {\sl XMM-Newton/NuSTAR} and 2017 {\sl NuSTAR} observations of Mrk\,110. (t) means that the parameter has been tied between the three epochs. 
}
\centering                          
\begin{tabular}{@{}l l l l}       
\hline\hline                 
 Parameters               &   \multicolumn{1}{c}{2019 Nov}   &   \multicolumn{1}{c}{2020 Apr} &   \multicolumn{1}{c}{2017 Jan} \\
                          &   \multicolumn{1}{c}{\sl NuSTAR} & \multicolumn{1}{c}{\sl NuSTAR} &   \multicolumn{1}{c}{\sl NuSTAR}\\
                          &   \multicolumn{1}{c}{+{\sl XMM}} & \multicolumn{1}{c}{+{\sl XMM}} & \\
 
 \hline
\hline   
& \multicolumn{3}{c}{{\sc zcutoffpl + zga}}\\
\hline
\hline
$E_{\rm cut}$ (keV) & 187$^{+100}_{-51}$  & 216$^{+157}_{-64}$ & 148$^{+34}_{-19}$   \\
$\Gamma$ &  1.74$\pm$0.01 &  1.69$\pm$0.02   &  1.72$\pm$0.01 \\
$norm_{\rm zcutoffpl}$ ($\times$10$^{-3}$) & 7.0$\pm$0.2  & 5.7$\pm$0.1 & 10.0$\pm$0.2 \\
$E_{\rm line}$ (keV) &  \multicolumn{3}{c}{6.37$^{+0.03}_{-0.04}$ (t)}\\
$\sigma_{\rm line}$ (keV) &\multicolumn{3}{c}{0.22$^{+0.07}_{-0.04}$ (t)}\\
$norm_{\rm line}$ ($\times$10$^{-5}$) & 2.5$\pm$0.4 & 2.0$^{+0.3}_{-0.4}$ & 3.0$\pm$0.4\\
$\chi^{2}$/d.o.f.  ($\chi^2_{\rm red}$) & \multicolumn{3}{c}{1801.6/1817  (0.99)}\\
\hline    
\hline                  
& \multicolumn{3}{c}{{\sc relxill}}\\
\hline
\hline
$E_{\rm cut}$ (keV) & 113$^{+28}_{-21}$  & 126$^{+35}_{-26}$ & 117$^{+12}_{-17}$   \\
$\Gamma$ &  1.73$^{+0.03}_{-0.02}$&  1.69$^{+0.03}_{-0.02}$   &  1.74$^{+0.01}_{-0.02}$ \\
$R_{\rm in}$ ($R_{\rm g}$) & 25$^{+15}_{-7}$ & 24$^{+15}_{-8}$ & 45$^{+80}_{-18}$\\
log\,$\xi$ (erg\,cm\,s$^{-1}$)& 2.9$\pm$0.2  &  2.9$\pm$0.2  & 2.8$\pm$0.1  \\
$\cal{R}$ & 0.16$\pm$0.03 & 0.15$^{+0.04}_{-0.03}$ & 0.12$^{+0.03}_{-0.02}$ \\
$norm_{\rm relxill}$ ($\times$10$^{-4}$) & 1.3$\pm$0.1  & 1.3$\pm$0.1 & 2.1$\pm$0.1 \\
$\chi^{2}$/d.o.f.  ($\chi^2_{\rm red}$) & \multicolumn{3}{c}{1843.3/1813  (1.02)}  \\
\hline    
\hline                  
 & \multicolumn{3}{c}{{\sc relxillcp}}\\
\hline
\hline
$kT_{\rm hot}$ (keV) & 26$^{+8}_{-5}$  & 26$^{+8}_{-5}$   & 26$^{+4}_{-3}$  \\
$\Gamma$ & 1.82$^{+0.01}_{-0.02}$   &  1.77$^{+0.02}_{-0.01}$ & 1.81$\pm$0.01     \\
$R_{\rm in}$ ($R_{\rm g}$) & 23$^{+10}_{-6}$ & 23$^{+14}_{-7}$ & 37$^{+43}_{-14}$\\
log\,$\xi$ (erg\,cm\,s$^{-1}$)& 3.0$^{+0.2}_{-0.21}$  &  3.1$\pm$0.2 & 3.0$^{+0.1}_{-0.2}$  \\
$\cal{R}$ & 0.14$\pm$0.03 & 0.13$\pm$0.03 & 0.09$^{+0.02}_{-0.01}$ \\
$norm_{\rm relxillcp}$ ($\times$10$^{-4}$) & 1.3$\pm$0.1  & 1.2$\pm$0.1 & 2.0$\pm$0.1  \\
$\chi^{2}$/d.o.f.  ($\chi^2_{\rm red}$) & \multicolumn{3}{c}{1929.7/1813  (1.06)}  \\
\hline    
\hline                  
& \multicolumn{3}{c}{{\sc reflkerr} (slab geometry)}\\
\hline
\hline
$kT_{\rm hot}$ (keV)                 & 30$\pm$4                & 31$^{+5}_{-3}$         & 30$^{+2}_{-3}$  \\
$\tau_{\rm hot}$               & 2.2$\pm$0.2             & 2.2$\pm$0.1            & 2.2$\pm$0.1   \\
$R_{\rm in}$ ($R_{\rm g}$)           & 26$^{+10}_{-8}$         & 26$\pm$9               & 42$^{+7}_{-13}$\\
log\,$\xi$ (erg\,cm\,s$^{-1}$)       & 3.1$^{+0.1}_{-0.2}$     & 3.1$^{+0.1}_{-0.2}$    & 2.8$^{+0.2}_{-0.1}$    \\
$\cal{R}$                            & 0.22$^{+0.01}_{-0.03}$  & 0.21$\pm$0.02          & 0.17$^{+0.02}_{-0.03}$\\
$norm_{\rm reflkerr}$ ($\times$10$^{-3}$) & 6.4$\pm$0.2        & 5.3$^{+0.1}_{-0.2}$    & 10.1$^{+0.3}_{-0.2}$   \\
$\chi^{2}$/d.o.f.  ($\chi^2_{\rm red}$) & \multicolumn{3}{c}{1850.4/1813 (1.02)}  \\
\hline
\hline
& \multicolumn{3}{c}{{\sc reflkerr} (spherical geometry)}\\
\hline
\hline
$kT_{\rm hot}$ (keV)                     & 22$\pm$1     & 21$\pm$1                      & 22$\pm$1  \\
$\tau_{\rm hot}$                   & 6.1$^{+0.2}_{-0.1}$   & 6.9$^{+0.1}_{-0.7}$  & 5.9$^{+0.4}_{-0.1}$   \\
$R_{\rm in}$ ($R_{\rm g}$)               & 25$^{+10}_{-7}$    & 22$^{+20}_{-6}$         & 31$^{+19}_{-6}$ \\
log\,$\xi$ (erg\,cm\,s$^{-1}$)           & 3.1$\pm$0.1        & 3.4$\pm$0.1            & 2.8$\pm$0.1 \\
$\cal{R}$                                & 0.12$\pm$0.02      & 0.16$^{+0.05}_{-0.03}$&0.09$^{+0.01}_{-0.02}$\\
$norm_{\rm reflkerr}$ ($\times$10$^{-3}$)& 6.7$^{+0.1}_{-0.3}$& 4.8$^{+0.3}_{-0.5}$  & 10.7$\pm$0.1 \\
$\chi^{2}$/d.o.f.  ($\chi^2_{\rm red}$) & \multicolumn{3}{c}{1913.5/1813 (1.05)}  \\
\hline    
\hline                  
$F^{\rm unabs}_{\rm 3-79\,keV}$ (erg\,cm$^{-2}$\,s$^{-1}$) &  6.9$\times$10$^{-11}$ & 6.2$\times$10$^{-11}$& 9.5$\times$10$^{-11}$  \\
$L^{\rm unabs}_{\rm 3-79\,keV}$ (erg\,s$^{-1}$) & 2.0$\times$10$^{44}$ & 1.8$\times$10$^{44}$ & 2.7$\times$10$^{44}$\\
\hline    
\hline                  
\end{tabular}
\label{tab:above3keV}
\end{table}

\begin{figure}[t!]
\begin{tabular}{c}
\includegraphics[width=0.9\columnwidth,angle=0]{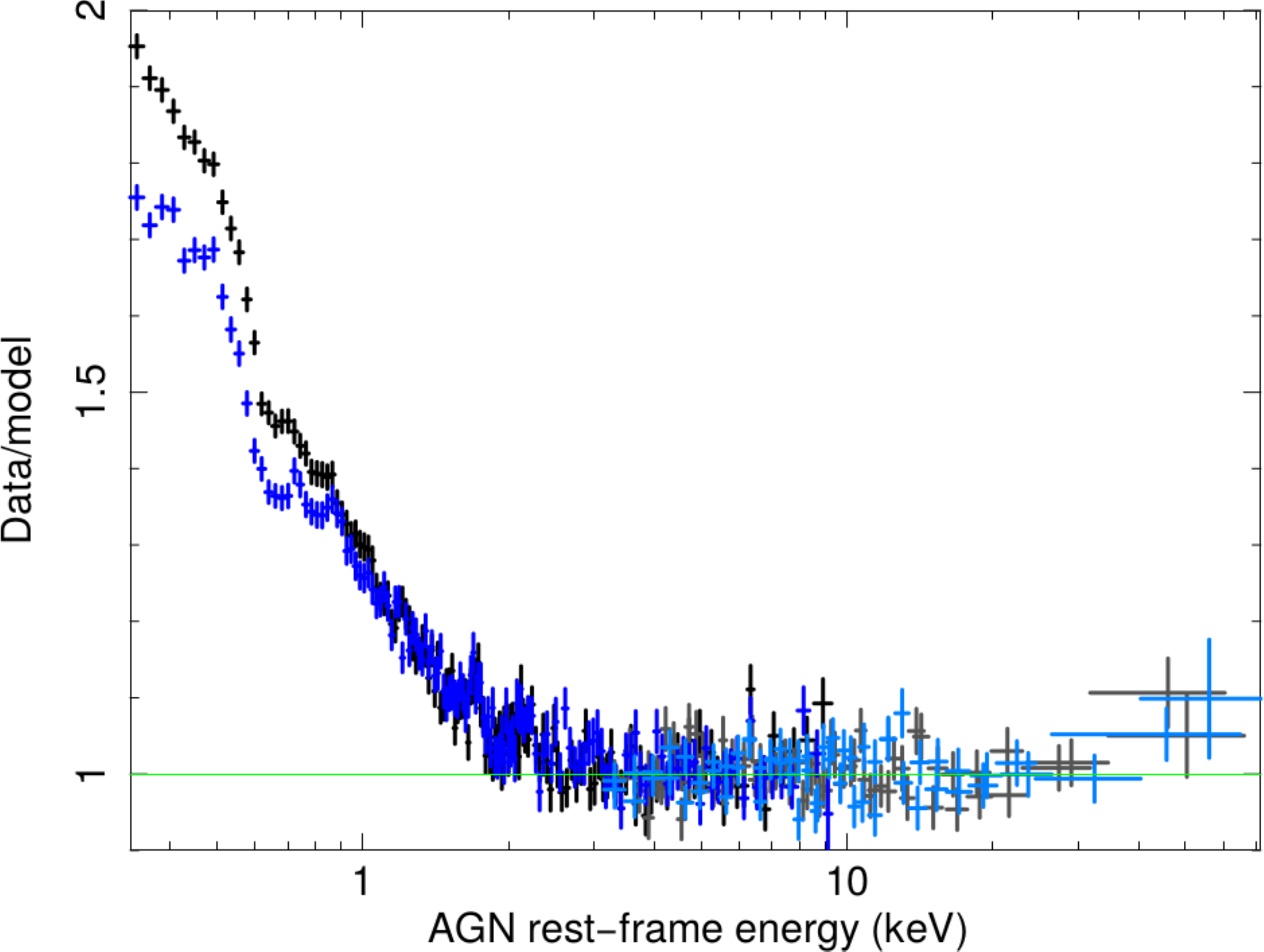} \\
\includegraphics[width=0.9\columnwidth,angle=0]{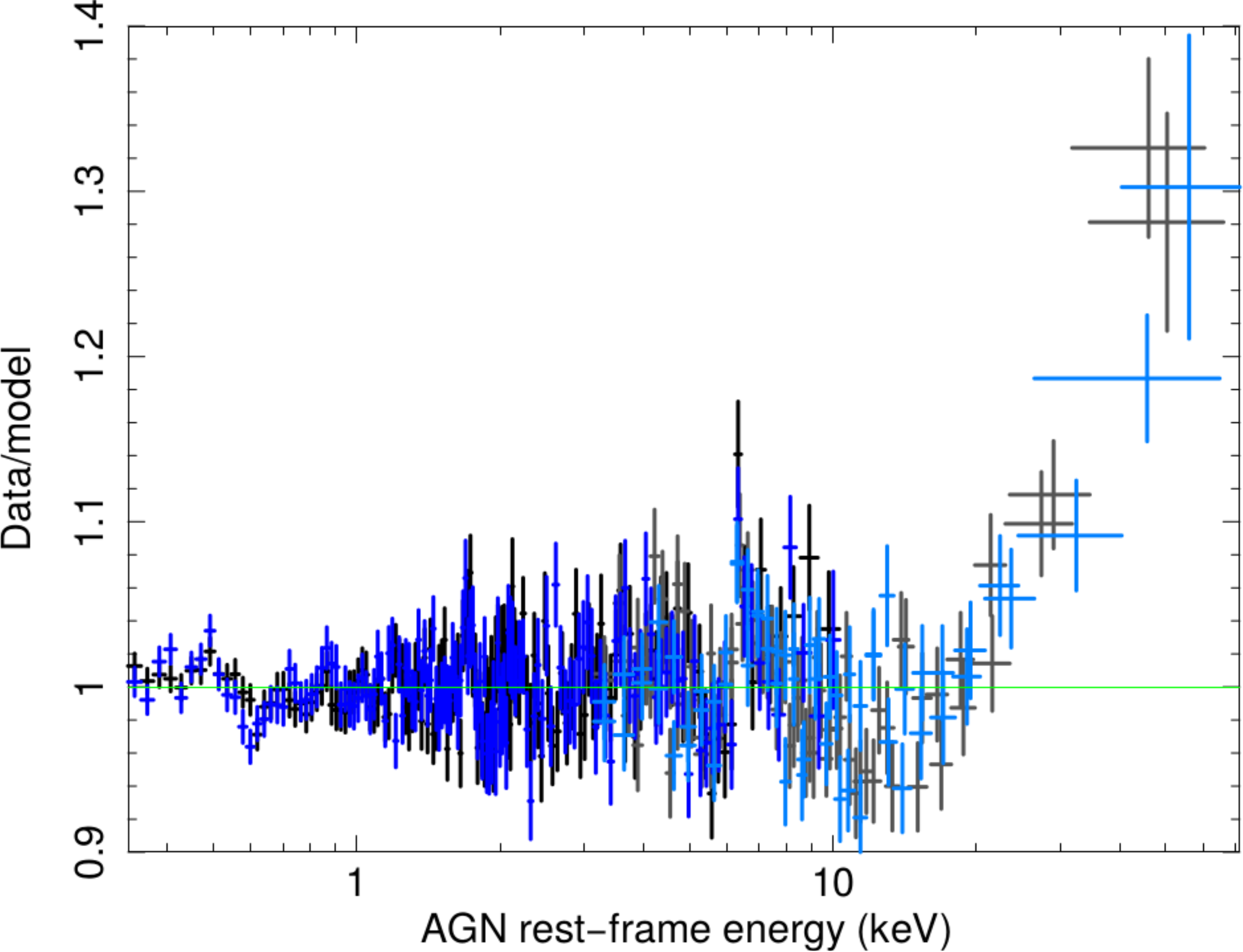}\\
\end{tabular}
\caption{Fits of the two 2019 and 2020 simultaneous {\sl XMM-Newton}/pn and {\sl NuSTAR} spectra with 
	the {\sc relxill} relativistic reflection model that assumes a coronal geometry for the disc emissivity.
Black: 2019 {\sl XMM-Newton}/pn, dark grey: 2019 {\sl NuSTAR}, 
 blue: 2020 {\sl XMM-Newton}/pn, light blue: 2020 {\sl NuSTAR}.  
	{\it Top panel:} Fit above 3\,keV, which has been extrapolated down to 0.3\,keV.
        The fit parameters are reported in Table~\ref{tab:above3keV}. 
        {\it Bottom panel}: Fit over the full 0.3--79\,keV energy range.  
        The fit parameters are reported in Table~\ref{tab:refl}. 
} 
\label{fig:refl}
\end{figure}

The {\sc relxill} model allows for relativistic reflection assuming a primary power law ($\Gamma$) continuum with an exponential cut-off at high energy ($E_{\rm cut}$). A good fit is found (Table~\ref{tab:above3keV}). 
The inner radii, where the reflection occurs, are inferred to be a few tens of $R_{\rm g}$. 
Moreover, the inferred very small reflection fraction values (ratio of intensity emitted towards the disc compared to that escaping to infinity) of $\mathcal{R}$\,$\simeq$\,0.1--0.2, and the high energy cut-off values of $E_{\rm cut}$\,$\sim$110--130\,keV for the three  epochs indicate that the lack of Compton hump is due to the weak reflection strength combined 
 with the presence of a moderately low high-energy cut-off value. 
 
 We now investigate relativistic reflection models with a primary Comptonisation continuum shape, which is more physical and has a sharper high-energy roll-over compared to an exponential cut-off power-law. In addition, such models have the advantage of having the hot corona temperature ($kT_{\rm hot}$) as a physical parameter rather than a phenomenological exponential cut-off energy. 
We first apply the {\sc relxillcp} model, which uses the {\sc nthcomp} Comptonisation model \citep{Zdziarski96,Zycki99} as the incident spectrum. 
The other physical parameters are the same as those in the {\sc relxill} model presented above. 
We find a good fit and infer similar hot corona temperatures of $kT_{\rm hot}$\,$\sim$26\,keV for all three epochs (Table~\ref{tab:above3keV}). 
Then, we consider the {\sc reflkerr} where the hard X-ray Comptonisation spectrum is computed with the {\sc compps} model \citep{Poutanen96}, which appears to be a better description of thermal Comptonisation when compared to Monte Carlo simulations \citep{Zdziarski20}. Moreover, {\sc reflkerr} has, as physical parameter, either the Compton parameter ($y$) or the optical depth ($\tau$). We choose to perform the fit with the optical depth as the direct inferred parameter. The temperature of the thermal seed photons ($kT_{\rm bb}$) Comptonised by the hot corona is an explicit physical parameter of this model. Here, we assume that the seed photons are provided by the cold disc, then $kT_{\rm bb}$ is fixed at 10\,eV corresponding to the expected maximum temperature of the accretion disc around a black hole mass of 1.4$\times$10$^{8}$\,M$_{\odot}$ accreting at a $\sim$10\% Eddington rate. In addition, {\sc reflkerr} allows us to choose either a slab or a spherical geometry for the hot corona. The latter corresponds to numerous active sphere regions above the disc surface. The hard X-ray shape of the reflected component is calculated using {\sc ireflect} convolved with {\sc compps} rather than using {\sc xillvercp} (see \citealt{Niedzwiecki19} for detailed explanations). Both models give good fits (Table~\ref{tab:above3keV}, Fig.~\ref{fig:reflabove3keV}), albeit a larger $\chi^2$ value for the spherical geometry. Similar values of the hot corona temperature are measured for the three epochs: $kT_{\rm hot}$\,$\sim$30--31\,keV and $kT_{\rm hot}$\,$\sim$21--22\,keV for the slab and spherical geometries, respectively. 
From $\tau$, we can infer the corresponding Compton-parameter of the hot corona ($y_{\rm hot}$) using the relation $y$=4$\tau$($kT$/511\,keV) \citep{Beloborodov99}. This correspond to $y_{\rm hot}$$\sim$0.5 and $y_{\rm hot}$$\sim$1.1 for the slab and spherical geometries, respectively. \\
 
The extrapolation of the fits down to 0.3\,keV for the 2019 and 2020 observations shows that the soft excess is not accounted for by any of these models, leaving a huge positive residual below 2\,keV. Fig.~\ref{fig:refl} (top panel) displays the extrapolation of the fit with the {\sc relxill} model for illustration purposes. 
We notice that over the 0.3--10\,keV energy range, the {\sc relxill} model provides a good fit (Table~\ref{tab:refl}, Fig.~\ref{fig:reflbelow10keV}), while when this model is applied to the whole 0.3--79\,keV X-ray broad-band range, it fails to both reproduce the soft and hard X-ray shapes (Fig.~\ref{fig:refl}, bottom panel).  
Similar results are found for other relativistic models (see appendix~\ref{sec:appendix} for more details).


\section{The X-ray broad band analysis of the 2019 and 2020 observations: A mixed Comptonisation and mild relativistic reflection modelling}\label{sec:Compt/rel}

We now aim to reproduce the X-ray broad-band spectra of Mrk\,110 with a model 
where Comptonisation is the dominant process combined with a moderate relativistic reflection component 
($\S$\ref{sec:main}; \citealt{Reeves21b}). 
Indeed, as shown above, relativistic reflection alone is not able to explain the soft X-ray excess, 
thereby we investigate the alternative scenario where it (mainly) originates from a warm corona. 

\begin{figure}[t!]
\begin{tabular}{c}
\includegraphics[width=0.9\columnwidth,angle=0]{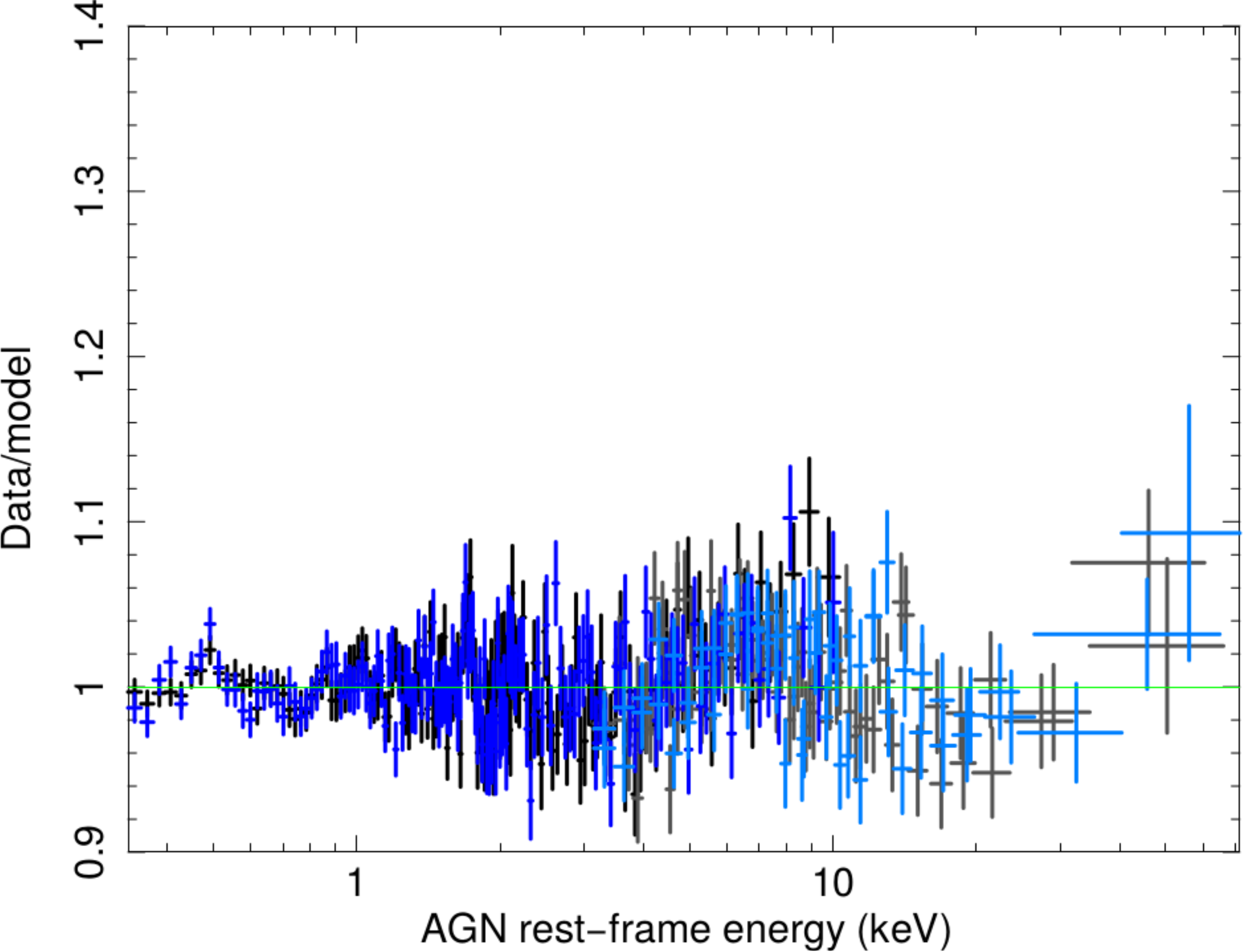}\\
\includegraphics[width=0.9\columnwidth,angle=0]{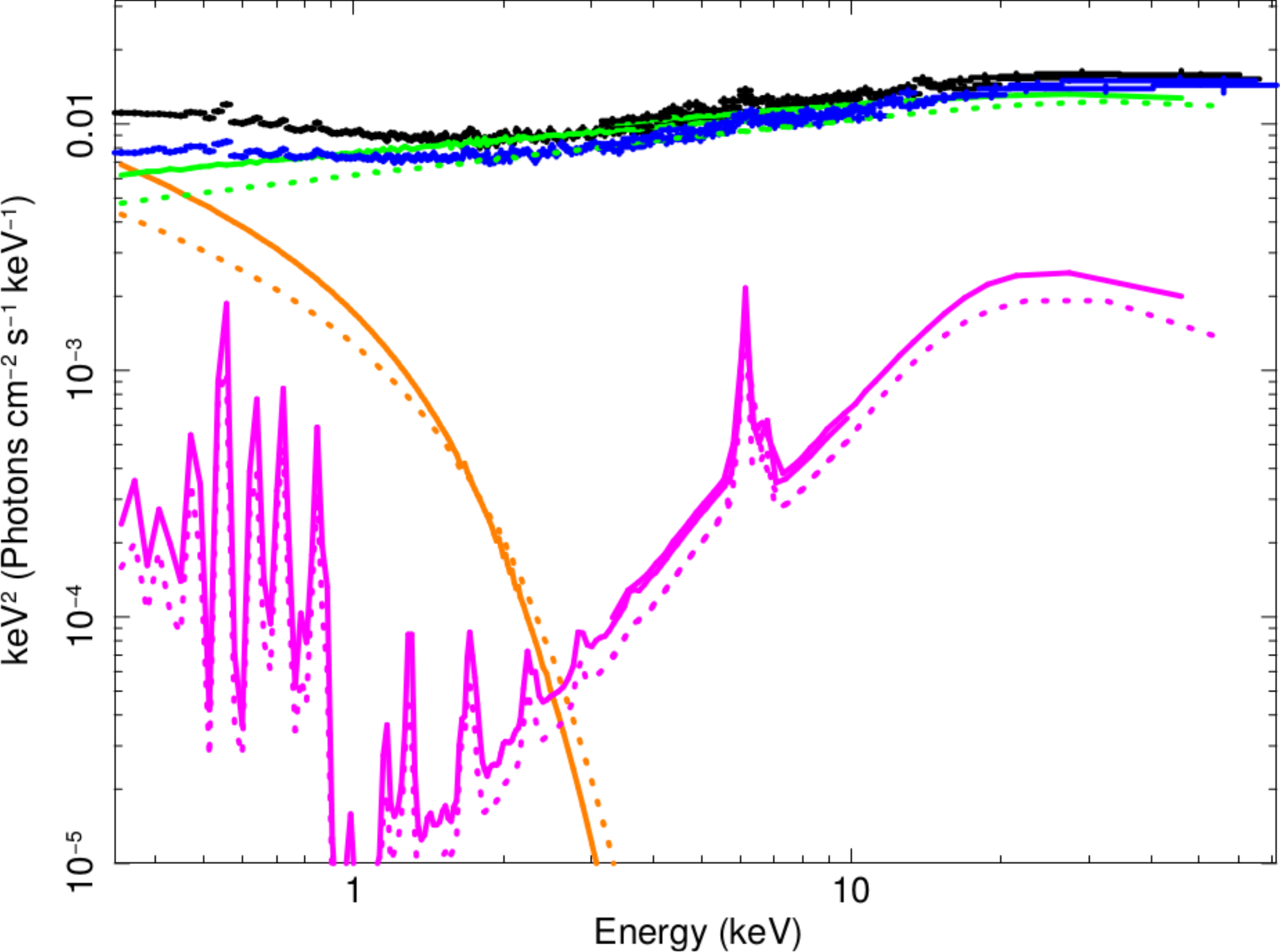}\\
\end{tabular}
\caption{Fit of the two simultaneous 2019 and 2020 {\sl XMM-Newton}/pn and {\sl NuSTAR} spectra 
with the {\sc tbnew(Gal)$\times$[comptt(warm)~+~reflkerr(hot)~+~reflkerr(refl)]} model.
The fit parameters are reported in Table~\ref{tab:Compt/rel}.
{\it Top panel}: Data/model ratio. 
Black: 2019 {\sl XMM-Newton}/pn, dark grey: 2019 {\sl NuSTAR}, 
 blue: 2020 {\sl XMM-Newton}/pn, light blue: 2020 {\sl NuSTAR}. 
The Y-axis range is identical to that of Fig.~\ref{fig:refl} to allow for a direct comparison. 
{\it Bottom panel}: Unfolded unabsorbed spectra (black: 2019, blue: 2020) where the separate contributions of the model components are
displayed (solid and dotted curves for 2019 and 2020, respectively): 
Orange: soft Comptonisation (warm corona); green: hard Comptonisation (hot corona); and magenta: disc reflection
 component.}  
\label{fig:Compt/rel}
\end{figure}

We apply to the two simultaneous {\sl XMM-Newton} and {\sl NuSTAR} spectra the following model: 
{\sc tbnew(Gal)$\times$[comptt(warm corona)~+~reflkerr(hot corona)~+~reflkerr(reflection)]}.  
 Since the {\sc compps} model cannot be used for coronal temperature value 
of a few 100s of keV, we instead make use of the
{\sc comptt} model to take for the warm corona contribution and 
infer its temperature, $kT_{\rm warm}$, and optical depth, $\tau_{\rm warm}$ \citep{Titarchuk94}. 
 The {\sc reflkerr} model is split into two components in order to disentangle 
 and display the contributions of the hot corona and of the moderate relativistic reflection. 
 For the former, the reflection strength is set to zero to only calculate the emission from the hot corona, while for the latter it is set to a negative value (here set to -0.2; $\S$\ref{sec:above3keV}) in order to only return the reflection component due to the illumination of the accretion disc by the hot corona. 
 As in $\S$\ref{sec:above3keV}, we fix the seed photon temperature ($kT_{\rm bb}$) at 10\,eV corresponding to the maximum temperature expected from the cold disc. A slab geometry for the hot corona is assumed.  
The other parameters of the two {\sc reflkerr} components are tied together for each epoch; 
we also set the emissivity index to the standard value of three and  
  allow the inner radius, where the relativistic reflection occurs, to be free to vary. 
   The hot corona temperature values have been fixed at those found in section~\ref{sec:above3keV}. 
The disc inclination degree is still set to 9.9\,degrees. 
The spin value is not constrained and then fixed at zero, 
but we checked that taking other values has no significant impact on other inferred parameter values.
Since we previously found that the contribution of the molecular torus is negligible, we do not include it. 
We obtain an overall very good fit for the full X-ray broad-band energy range 
for both epochs (Fig.~\ref{fig:Compt/rel} top panel; Table~\ref{tab:Compt/rel}). 
The inferred ionization parameter, $\log\xi$=1.1$\pm$0.1, for the accretion disc agrees very well    
with that found for the \ion{O}{vii} soft X-ray emission line ($\log\xi$\,$\sim$1.2; \citealt{Reeves21b}), and 
is also consistent with the emission of the moderately broad neutral Fe\,K$\alpha$ emission line too 
(Fig.~\ref{fig:Compt/rel}, bottom panel, magenta curves). 
We derive warm Comptonised plasma temperature values of $kT_{\rm warm}$\,$\sim$\,0.3\,keV and high optical depth values of $\tau_{\rm warm}$\,$\sim$13 for the two epochs (Table~\ref{tab:Compt/rel}).
The variability of the soft X-ray excess between both epochs  
appears to be driven by its flux variability, being stronger when brighter (at 5.5$\sigma$ confidence level for the normalisation)
rather than by any statistical change of its physical parameters. 

\begin{table}[t!]
\caption{Simultaneous fits over the broad-band energy range 
of the two simultaneous {\sl XMM-Newton}/pn and {\sl NuSTAR} spectra  
 with {\sc tbnew(Gal)$\times$[comptt(warm)~+~reflkerr(hot)~+~reflkerr(refl)]}.
 See text for detailed explanations.
  The hot corona temperature values have been fixed at those found in section~\ref{sec:above3keV}, 
 and the temperature of the seed photons at 10\,eV.  
(t) means that the value has been tied between both epochs. 
(a) The 0.3--79\,keV unabsorbed fluxes are expressed in units of erg\,cm$^{-2}$\,s$^{-1}$.
(b) The 0.3--79\,keV intrinsic luminosities are expressed in units of erg\,s$^{-1}$.
}                          
\centering                          
\begin{tabular}{@{}l l l}       
\hline\hline                 
                &   \multicolumn{1}{c}{2019 Nov}   &   \multicolumn{1}{c}{2020 April} \\
\hline                                  
$kT_{\rm warm}$ (keV)& 0.26$\pm$0.02 &  0.30$\pm$0.03 \\
$\tau_{\rm warm}$ &  13.8$^{+0.8}_{-1.0}$ & 12.9$^{+1.0}_{-1.2}$ \\
$norm_{\rm comptt(warm})$ & 4.0$^{+0.2}_{-0.3}$ &  2.0$^{+0.2}_{-0.1}$\\
\hline
$kT_{\rm hot}$ (keV)&  30 (f) & 31 (f)  \\
$\tau_{\rm hot}$ &  2.1$\pm$0.2 & 2.1$\pm$0.1 \\
$norm_{\rm reflkerr(hot)}$ & 7.6$\pm$0.1$\times$10$^{-3}$ & 6.2$\pm$0.1$\times$10$^{-3}$\\
\hline
log\,$\xi$ (erg\,cm\,s$^{-1}$) & \multicolumn{2}{c}{1.1$\pm$0.1 (t)}       \\
$R_{\rm in}$ ($R_{\rm g}$) & $\geq$58 & $\geq$35\\
$\cal{R}$ & -0.2 (f) & -0.2 (f)\\
$norm_{\rm reflkerr(refl)}$ & 1.3$\pm$0.2$\times$10$^{-2}$  & 0.9$\pm$0.2$\times$10$^{-2}$\\
\hline
$F^{\rm unabs}_{\rm 0.3-79\,keV}$$^{(a)}$ & 11.0$\times$10$^{-11}$ &  9.1$\times$10$^{-11}$ \\
$L_{\rm 0.3-79\,keV}$$^{(b)}$ & 3.2$\times$10$^{44}$&  2.6$\times$10$^{44}$\\
\hline           
$\chi^{2}$/d.o.f.  & \multicolumn{2}{c}{1755.5/1620}  \\
$\chi^{2}_{\rm red}$ & \multicolumn{2}{c}{1.08} \\
\hline    \hline                  
\end{tabular}
\label{tab:Compt/rel}
\end{table}

\begin{figure}[t!]
\begin{tabular}{c}
\includegraphics[width=0.9\columnwidth,angle=0]{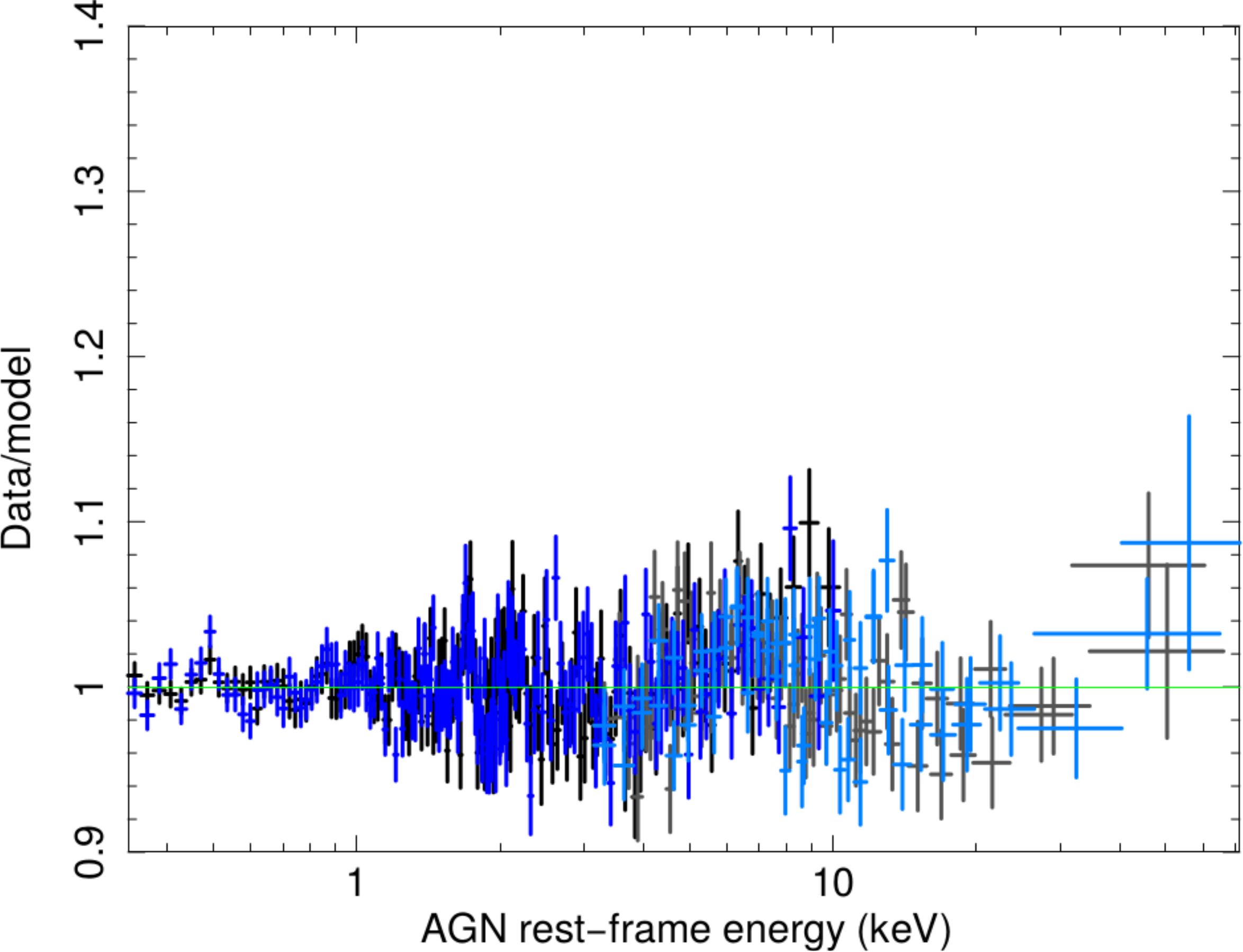}
\end{tabular}
\caption{Same as Fig.~\ref{fig:Compt/rel} (top panel) but with the seed photon temperature allowed to be free to vary. }  
\label{fig:Compt/rel2}
\end{figure}

Since there is a significant soft X-ray excess mainly due to the warm corona, these soft X-ray photons may be Comptonised by the hot corona too, as was found for \object{PKS 0558-504} \citep{Gliozzi13}. Therefore, 
 we allow the input seed photon temperature, $kT_{\rm bb}$, to vary, but noting that it is tied between the 2019 and 2020 epochs due to the lack of significant variability of the temperature of the soft excess. The ionization parameter of the accretion disc  is fixed at log\,$\xi$=1.1, as was found above. 
 The resulting seed photon temperature is $kT_{\rm bb}$=75$\pm$4\,eV, which is much higher than the maximum temperature of an accretion disc around a supermassive black hole of mass $\sim$10$^{8}$\,M$_{\odot}$ and instead corresponds to a black hole mass of about three decades lower and accreting near Eddington. 
This value could instead represent an intermediate temperature between the seed photons from the cold disc and from the warm corona. 
Therefore, the seed photons Comptonised by the electrons of the hot corona may be provided by a combination of both the cold disc and the warm corona. This is strengthened by the correlation between both the hard and soft X-ray fluxes and the X-ray and UV fluxes inferred from the long-term {\it Swift} monitoring of Mrk\,110 (Lobban et al.\ 2021, in prep.).

\section{Summary and discussion}\label{sec:discussion}

This paper presents the X-ray spectral analysis of two first simultaneous {\sl XMM-Newton} and {\sl NuSTAR} observations 
of Mrk\,110 performed within about a half-year interval on 2019 November 16--17 and in 2020 April 5--6. \\

A prominent and absorption-free smooth soft X-ray excess is observed in both observations, 
with the strongest one seen during the brightest observation (2019). 
This confirmed the presence of such a feature reported for the 2004 {\sl XMM-Newton} observation 
that caught the source in a slightly higher-flux state \citep{Boller07,Boissay16,JiangJ19,Gliozzi20}. 
A very similar spectral shape is found above 2--3\,keV for both the continuum 
and the Fe\,K$\alpha$ emission line. 

 The AGN rest-frame peak energy of the Fe\,K$\alpha$ line is slightly redshifted toward 6.4\,keV ($E$=6.37$^{+0.03}_{-0.04}$\,keV) and its width  (FWHM=22\,000$^{+7800}_{-4400}$\,km\,s$^{-1}$) indicates that the Fe\,K$\alpha$ line could originate from the accretion disc, as inferred for the \ion{O}{vii} soft X-ray emission line with the RGS data analysis \citep{Reeves21b}.
 The Fe\,K$\alpha$ profile is mainly produced by moderately broadened reflection arising from an almost face-on accretion disc. 
 Indeed, only a very weak possible additional contribution from the molecular torus ($EW_{\rm n}$\,$\lesssim$\,20\,eV) is found, 
as expected from the very small total luminosity of the torus measured by \cite{Ezhikode17} 
which is one of the lowest found in their sample. 
Only much higher spectral resolution near 6.4\,keV -- as targeted by the future, calorimeter resolution, 
X-ray missions {\sl XRISM} \citep{XRISM20} and {\sl ATHENA}  \citep{Barret18} -- 
will provide an accurate determination of the genuine possible contributions of the BLR and torus regions. 

The photon indices are found to be $\sim$1.7--1.8, confirming previous
measurements found with the 2004 {\sl XMM-Newton} \citep{Boller07}, 
the 2007 {\sl Suzaku} \citep{Patrick12,Walton13,Mantovani16,Waddel20}, and the 2017 {\sl NuSTAR} \citep{Ezhikode20,Panagiotou20} observations. 
Such values are much flatter than what are commonly observed in NLS1s, but are consistent with BLS1s \cite[e.g.,][]{P04a,Zhou10b,Waddel20,Gliozzi20}.  This is in agreement with the BLS1 classification of Mrk\,110 (see $\S$\ref{sec:Introduction}), rather than a NLS1 one which is still commonly assumed in the literature. 
Moreover, no Compton hump is observed in the {\sl NuSTAR} spectra, with instead a flat spectrum and  
a hard X-ray deficit above $\sim$30\,keV, as was found previously for the deep January 2017 {\sl NuSTAR} observation, 
which caught the source in a slightly higher flux level \citep{Panagiotou20,Ezhikode20}. \\

The spectral analysis above 3\,keV allows us to characterise the hot corona and the relativistic reflection emission properties. The    
  deep January 2017 {\sl NuSTAR} observation was also included in the simultaneous fit \citep{Ezhikode20,Panagiotou20}. 
From the {\sc relxill} reflection model, assuming a primary exponential cut-off power-law continuum, we find moderate reflection strengths, $\cal{R}$\,$\sim$0.1--0.2, and high cut-off energies at $E_{\rm cut}$\,$\sim$110--120\,keV.
These values are in very good agreement with those measured from the average long-term Swift BAT spectrum \citep{Vincentelli21}. 
Applying relativistic reflection models assuming a primary Comptonisation continuum, we infer the hot corona temperature to be $kT_{\rm hot}$\,$\sim$26--31\,keV ($kT_{\rm hot}$\,$\sim$21--22\,keV) and the optical depth to be $\tau_{\rm hot}$\,$\sim$2  ($\tau_{\rm  hot}$\,$\sim$6--7) for the slab (or spherical) geometry.  
From the spectral analysis, it is not possible to discriminate between either of the hot corona geometries, although the slab geometry provides a better fit. In the near future, X-ray polarimetry is expected to play an important role in such a framework \cite[e.g.,][]{Schnittman10,Beheshtipour17,Tamborra18,Marinucci19}, thanks to, for example, {\sl IXPE} \citep{Weisskopf16} and {\sl eXTP} \citep{ZhangS16}. 
 While the corona temperatures found for Mrk\,110 are broadly consistent with the average ones found 
by \cite{Middei19} from a 
sample of 26 AGN (with <$kT_{\rm hot}$>=50$\pm$21\,keV and <$kT_{\rm hot}$>=53$\pm$23\,keV for the slab and spherical geometry, respectively), it is likely to be located in the lower range of this distribution.
 However, its hot coronal temperature is not as low as 
the temperatures inferred for some AGN with much lower high-energy cut-offs, 
such as  \object{GRS 1734-292} \citep{Tortosa17}, \object{Ark\,564} \citep{Kara17}, and \object{PDS\,456} \citep{Reeves21a}, where $kT$ could be as low as 15\,keV.  

The lack of variability of the cut-off energy and of the hot corona temperature 
may be due to the fact that during the 2017, 2019 and 2020 {\sl NuSTAR} 
observations the source was caught in a very similar hard X-ray flux state
with a 3--79\,keV flux variability of only 1.5 between the 2017 and 2020 observations. 
Therefore, any possible `hotter/cooler-when-brighter' 
behavior cannot be confirmed here \citep{Keek16,Ursini16,ZhangJ18,Middei19,Kang21}. 
Further broad-band X-ray monitoring, using the unique synergy of {\sl XMM-Newton} and {\sl NuSTAR}, 
of this bright source, is mandatory for any definitive statement 
 about the physical and/or geometrical changes of the hot corona. 
 In particular, the observations would need to sample a larger flux range, as is observed 
in the long-term {\sl Swift} monitoring of Mrk\,110 \citep[][Lobban et al.\ 2021, in prep.]{Vincentelli21}. \\

The X-ray broad-band analysis shows that relativistic reflection models due to illumination on a constant density accretion disc alone cannot explain the X-ray broad-band {\sl XMM-Newton/NuSTAR} spectra of Mrk\,110 (appendix~\ref{sec:appendix}); while, 
data below 10\,keV does not allow us to rule out pure relativistic reflection models (Fig.~\ref{fig:reflbelow10keV}). 
 This confirms that X-ray broad-band spectra are mandatory to determine the physical processes at work
in disc-corona systems in AGN and their relative contributions \citep{Matt14,Porquet18,Middei20,Ursini20,Walton20}. 
Instead, the broad-band energy analysis demonstrates that warm and hot Comptonisation are the dominant processes, 
though a moderate relativistic reflection component is also present. 
The latter occurs at distance of a few 10s of $R_{\rm g}$ (see $\S$\ref{sec:above3keV} and \citealt{Reeves21b}) and is responsible for both the moderately broad \ion{O}{vii} and  Fe\,K$\alpha$ emission lines.  
We also find that both the cold disc and the warm corona can provide seed photons to the hot corona. 
  Such a model combining Comptonisation and relativistic reflection has been proven to explain well the X-ray broad-band 
 spectra of several other AGN, such as: \object{Ark\,120} \citep{Porquet18}, \object{Fairall\,9} \citep{Lohfink16}, 
 \object{Ton\,S180} \citep{Matzeu20}, and \object{ESO\,362-G18} \citep{XuY21}. 
  In a forthcoming article, we will study the spectral energy distribution of Mrk\,110 from optical/UV to hard X-ray excess, as done previously done for the bare AGN Ark\,120 \citep{Porquet19}. This will allows us to take both the geometry and energetic of the warm/hot corona and the outer disc into account, to determine the accretion rate and to measure the black hole spin \citep[e.g.,][]{Done12,Kubota18}.\\

 The physical properties of the warm corona ($kT_{\rm warm}$\,$\sim$0.3\,keV, $\tau_{\rm warm}$\,$\sim$13) 
 are compatible during both the 2019 and 2020 XMM/NuSTAR observations of Mrk\,110, with only the warm corona flux  
  found to be significantly different (at $\sim$5.5$\sigma$ confidence level). 
 They are also consistent with those measured ($kT_{\rm warm}$\,$\sim$0.3--1\,keV, $\tau_{\rm warm}$\,$\sim$10--20) in {\sl XMM-Newton} AGN samples \citep[e.g.,][]{P04a,Bianchi09}, and in several individual X-ray broad-band spectral analysis of AGN, such as: \object{Mrk\,509} \citep{Petrucci13,Mehdipour15}, \object{Ark\,120} \citep{Matt14,Porquet18}, \object{HE 1143-1810} \citep{Ursini20}, \object{NGC 4593} \citep{Middei19}, \object{Mrk 359} \citep{Middei20}, \object{TON\,S180} \citep{Matzeu20}, and \object{ESO\,511-G030} \citep{Ghosh21}. 
 Very recent state-of-the-art simulations have been performed showing 
 that a warm corona with such properties can indeed exist,
 until sufficient internal mechanical heating is present, and that warm and hot coronae can co-exist \citep{Petrucci18,Petrucci20,Ballantyne20a,Ballantyne20b}.  
 Moreover, these authors found that no strong absorption/emission lines are formed 
 which can well explain the absorption-free smooth soft X-ray shape
 observed in bare AGN, or in most other AGN when the warm absorber and any 
 moderate relativistic reflection contributions have been removed. \\

\begin{acknowledgements}
  
D.\ Porquet dedicates this paper to her friend and colleague, C\'{e}cile Renault, who
has  sadly  passed  away on April 2021.  
The authors thank the anonymous referee for useful
 and constructive comments. 
The paper is based on observations obtained with the {\sl XMM-Newton}, and ESA science
mission with instruments and contributions directly funded by ESA
member states and the USA (NASA). 
 This work made use of data from the
{\sl NuSTAR} mission, a project led by the California Institute of
Technology, managed by the Jet Propulsion Laboratory, and
funded by NASA. 
This research has made use of the {\sl NuSTAR} 
Data Analysis Software (NuSTARDAS) jointly developed by
the ASI Science Data Center and the California Institute of
Technology.
This work was supported by CNES. 
\end{acknowledgements}

%
%

\bibliographystyle{aa}
\bibliography{biblio}

\begin{thebibliography}{95}
\expandafter\ifx\csname natexlab\endcsname\relax\def\natexlab#1{#1}\fi

\bibitem[{{Afanasiev} {et~al.}(2019){Afanasiev}, {Popovi{\'c}}, \&
  {Shapovalova}}]{Afanasiev19}
{Afanasiev}, V.~L., {Popovi{\'c}}, L.~{\v{C}}., \& {Shapovalova}, A.~I. 2019,
  \mnras, 482, 4985

\bibitem[{{Arnaud}(1996)}]{Arnaud96}
{Arnaud}, K.~A. 1996, in ASP Conf. Ser. 101: Astronomical Data Analysis
  Software and Systems V, ed. G.~H. {Jacoby} \& J.~{Barnes}, 17

\bibitem[{{Ballantyne}(2020)}]{Ballantyne20a}
{Ballantyne}, D.~R. 2020, \mnras, 491, 3553

\bibitem[{{Ballantyne} \& {Xiang}(2020)}]{Ballantyne20b}
{Ballantyne}, D.~R. \& {Xiang}, X. 2020, \mnras, 496, 4255

\bibitem[{{Balokovi{\'c}} {et~al.}(2018){Balokovi{\'c}}, {Brightman},
  {Harrison}, {Comastri}, {Ricci}, {Buchner}, {Gandhi}, {Farrah}, \&
  {Stern}}]{Balokovic18}
{Balokovi{\'c}}, M., {Brightman}, M., {Harrison}, F.~A., {et~al.} 2018, \apj,
  854, 42

\bibitem[{{Balokovi{\'c}} {et~al.}(2019){Balokovi{\'c}}, {Garc{\'\i}a}, \&
  {Cabral}}]{Balokovic19}
{Balokovi{\'c}}, M., {Garc{\'\i}a}, J.~A., \& {Cabral}, S.~E. 2019, Research
  Notes of the American Astronomical Society, 3, 173

\bibitem[{{Barret} {et~al.}(2018){Barret}, {Lam Trong}, {den Herder}, {Piro},
  {Cappi}, {Houvelin}, {Kelley}, {Mas-Hesse}, {Mitsuda}, {Paltani}, {Rauw},
  {Rozanska}, {Wilms}, {Bandler}, {Barbera}, {Barcons}, {Bozzo}, {Ceballos},
  {Charles}, {Costantini}, {Decourchelle}, {den Hartog}, {Duband}, {Duval},
  {Fiore}, {Gatti}, {Goldwurm}, {Jackson}, {Jonker}, {Kilbourne}, {Macculi},
  {Mendez}, {Molendi}, {Orleanski}, {Pajot}, {Pointecouteau}, {Porter},
  {Pratt}, {Pr{\^e}le}, {Ravera}, {Sato}, {Schaye}, {Shinozaki}, {Thibert},
  {Valenziano}, {Valette}, {Vink}, {Webb}, {Wise}, {Yamasaki}, {Douchin},
  {Mesnager}, {Pontet}, {Pradines}, {Branduardi-Raymont}, {Bulbul}, {Dadina},
  {Ettori}, {Finoguenov}, {Fukazawa}, {Janiuk}, {Kaastra}, {Mazzotta},
  {Miller}, {Miniutti}, {Naze}, {Nicastro}, {Scioritino}, {Simonescu},
  {Torrejon}, {Frezouls}, {Geoffray}, {Peille}, {Aicardi}, {Andr{\'e}},
  {Daniel}, {Cl{\'e}net}, {Etcheverry}, {Gloaguen}, {Hervet}, {Jolly}, {Ledot},
  {Paillet}, {Schmisser}, {Vella}, {Damery}, {Boyce}, {Dipirro}, {Lotti},
  {Schwander}, {Smith}, {Van Leeuwen}, {van Weers}, {Clerc}, {Cobo}, {Dauser},
  {Kirsch}, {Cucchetti}, {Eckart}, {Ferrando}, \& {Natalucci}}]{Barret18}
{Barret}, D., {Lam Trong}, T., {den Herder}, J.-W., {et~al.} 2018, in Society
  of Photo-Optical Instrumentation Engineers (SPIE) Conference Series, Vol.
  10699, Space Telescopes and Instrumentation 2018: Ultraviolet to Gamma Ray,
  ed. J.-W.~A. {den Herder}, S.~{Nikzad}, \& K.~{Nakazawa}, 106991G

\bibitem[{{Beheshtipour} {et~al.}(2017){Beheshtipour}, {Krawczynski}, \&
  {Malzac}}]{Beheshtipour17}
{Beheshtipour}, B., {Krawczynski}, H., \& {Malzac}, J. 2017, \apj, 850, 14

\bibitem[{{Beloborodov}(1999)}]{Beloborodov99}
{Beloborodov}, A.~M. 1999, in Astronomical Society of the Pacific Conference
  Series, Vol. 161, High Energy Processes in Accreting Black Holes, ed.
  J.~{Poutanen} \& R.~{Svensson}, 295

\bibitem[{{Bian} \& {Zhao}(2002)}]{Bian02}
{Bian}, W. \& {Zhao}, Y. 2002, \aap, 395, 465

\bibitem[{{Bianchi} {et~al.}(2009){Bianchi}, {Guainazzi}, {Matt}, {Fonseca
  Bonilla}, \& {Ponti}}]{Bianchi09}
{Bianchi}, S., {Guainazzi}, M., {Matt}, G., {Fonseca Bonilla}, N., \& {Ponti},
  G. 2009, \aap, 495, 421

\bibitem[{{Bischoff} \& {Kollatschny}(1999)}]{Bischoff99}
{Bischoff}, K. \& {Kollatschny}, W. 1999, \aap, 345, 49

\bibitem[{{Boissay} {et~al.}(2016){Boissay}, {Ricci}, \& {Paltani}}]{Boissay16}
{Boissay}, R., {Ricci}, C., \& {Paltani}, S. 2016, \aap, 588, A70

\bibitem[{{Boller} {et~al.}(2007){Boller}, {Balestra}, \&
  {Kollatschny}}]{Boller07}
{Boller}, T., {Balestra}, I., \& {Kollatschny}, W. 2007, \aap, 465, 87

\bibitem[{{Boroson} \& {Green}(1992)}]{Boroson92}
{Boroson}, T.~A. \& {Green}, R.~F. 1992, \apjs, 80, 109

\bibitem[{{Crummy} {et~al.}(2006){Crummy}, {Fabian}, {Gallo}, \&
  {Ross}}]{Crummy06}
{Crummy}, J., {Fabian}, A.~C., {Gallo}, L., \& {Ross}, R.~R. 2006, \mnras, 365,
  1067

\bibitem[{{Dauser} {et~al.}(2013){Dauser}, {Garcia}, {Wilms}, {B{\"o}ck},
  {Brenneman}, {Falanga}, {Fukumura}, \& {Reynolds}}]{Dauser13}
{Dauser}, T., {Garcia}, J., {Wilms}, J., {et~al.} 2013, \mnras, 430, 1694

\bibitem[{{Dauser} {et~al.}(2010){Dauser}, {Wilms}, {Reynolds}, \&
  {Brenneman}}]{Dauser10}
{Dauser}, T., {Wilms}, J., {Reynolds}, C.~S., \& {Brenneman}, L.~W. 2010,
  \mnras, 409, 1534

\bibitem[{{Decarli} {et~al.}(2008){Decarli}, {Dotti}, {Fontana}, \&
  {Haardt}}]{Decarli08}
{Decarli}, R., {Dotti}, M., {Fontana}, M., \& {Haardt}, F. 2008, \mnras, 386,
  L15

\bibitem[{{Done} {et~al.}(2012){Done}, {Davis}, {Jin}, {Blaes}, \&
  {Ward}}]{Done12}
{Done}, C., {Davis}, S.~W., {Jin}, C., {Blaes}, O., \& {Ward}, M. 2012, \mnras,
  2196

\bibitem[{{Ezhikode} {et~al.}(2020){Ezhikode}, {Dewangan}, {Misra}, \&
  {Philip}}]{Ezhikode20}
{Ezhikode}, S.~H., {Dewangan}, G.~C., {Misra}, R., \& {Philip}, N.~S. 2020,
  \mnras, 495, 3373

\bibitem[{{Ezhikode} {et~al.}(2017){Ezhikode}, {Gandhi}, {Done}, {Ward},
  {Dewangan}, {Misra}, \& {Philip}}]{Ezhikode17}
{Ezhikode}, S.~H., {Gandhi}, P., {Done}, C., {et~al.} 2017, \mnras, 472, 3492

\bibitem[{{Fabian} {et~al.}(2012){Fabian}, {Zoghbi}, {Wilkins}, {Dwelly},
  {Uttley}, {Schartel}, {Miniutti}, {Gallo}, {Grupe}, {Komossa}, \&
  {Santos-Lle{\'o}}}]{Fabian12}
{Fabian}, A.~C., {Zoghbi}, A., {Wilkins}, D., {et~al.} 2012, \mnras, 419, 116

\bibitem[{{Fukazawa} {et~al.}(2011){Fukazawa}, {Hiragi}, {Mizuno}, {Nishino},
  {Hayashi}, {Yamasaki}, {Shirai}, {Takahashi}, \& {Ohno}}]{Fukazawa11}
{Fukazawa}, Y., {Hiragi}, K., {Mizuno}, M., {et~al.} 2011, \apj, 727, 19

\bibitem[{{F{\"u}rst} {et~al.}(2015){F{\"u}rst}, {Nowak}, {Tomsick}, {Miller},
  {Corbel}, {Bachetti}, {Boggs}, {Christensen}, {Craig}, {Fabian}, {Gandhi},
  {Grinberg}, {Hailey}, {Harrison}, {Kara}, {Kennea}, {Madsen}, {Pottschmidt},
  {Stern}, {Walton}, {Wilms}, \& {Zhang}}]{Fuerst15}
{F{\"u}rst}, F., {Nowak}, M.~A., {Tomsick}, J.~A., {et~al.} 2015, \apj, 808,
  122

\bibitem[{{Garc{\'{\i}}a} {et~al.}(2016){Garc{\'{\i}}a}, {Fabian}, {Kallman},
  {Dauser}, {Parker}, {McClintock}, {Steiner}, \& {Wilms}}]{Garcia16}
{Garc{\'{\i}}a}, J.~A., {Fabian}, A.~C., {Kallman}, T.~R., {et~al.} 2016,
  \mnras, 462, 751

\bibitem[{{Ghosh} \& {Laha}(2021)}]{Ghosh21}
{Ghosh}, R. \& {Laha}, S. 2021, \apj, 908, 198

\bibitem[{{Gliozzi} {et~al.}(2013){Gliozzi}, {Papadakis}, {Grupe}, {Brinkmann},
  \& {R{\"a}th}}]{Gliozzi13}
{Gliozzi}, M., {Papadakis}, I.~E., {Grupe}, D., {Brinkmann}, W.~P., \&
  {R{\"a}th}, C. 2013, \mnras, 433, 1709

\bibitem[{{Gliozzi} {et~al.}(2011){Gliozzi}, {Titarchuk}, {Satyapal}, {Price},
  \& {Jang}}]{Gliozzi11}
{Gliozzi}, M., {Titarchuk}, L., {Satyapal}, S., {Price}, D., \& {Jang}, I.
  2011, \apj, 735, 16

\bibitem[{{Gliozzi} \& {Williams}(2020)}]{Gliozzi20}
{Gliozzi}, M. \& {Williams}, J.~K. 2020, \mnras, 491, 532

\bibitem[{{Goodrich}(1989)}]{Goodrich89}
{Goodrich}, R.~W. 1989, \apj, 342, 224

\bibitem[{{Grupe}(2004)}]{Grupe04}
{Grupe}, D. 2004, \aj, 127, 1799

\bibitem[{{Harrison} {et~al.}(2013){Harrison}, {Craig}, {Christensen},
  {Hailey}, {Zhang}, {Boggs}, {Stern}, {Cook}, {Forster}, {Giommi},
  {Grefenstette}, {Kim}, {Kitaguchi}, {Koglin}, {Madsen}, {Mao}, {Miyasaka},
  {Mori}, {Perri}, {Pivovaroff}, {Puccetti}, {Rana}, {Westergaard}, {Willis},
  {Zoglauer}, {An}, {Bachetti}, {Barri{\`e}re}, {Bellm}, {Bhalerao},
  {Brejnholt}, {Fuerst}, {Liebe}, {Markwardt}, {Nynka}, {Vogel}, {Walton},
  {Wik}, {Alexander}, {Cominsky}, {Hornschemeier}, {Hornstrup}, {Kaspi},
  {Madejski}, {Matt}, {Molendi}, {Smith}, {Tomsick}, {Ajello}, {Ballantyne},
  {Balokovi{\'c}}, {Barret}, {Bauer}, {Blandford}, {Brandt}, {Brenneman},
  {Chiang}, {Chakrabarty}, {Chenevez}, {Comastri}, {Dufour}, {Elvis}, {Fabian},
  {Farrah}, {Fryer}, {Gotthelf}, {Grindlay}, {Helfand}, {Krivonos}, {Meier},
  {Miller}, {Natalucci}, {Ogle}, {Ofek}, {Ptak}, {Reynolds}, {Rigby},
  {Tagliaferri}, {Thorsett}, {Treister}, \& {Urry}}]{Harrison13}
{Harrison}, F.~A., {Craig}, W.~W., {Christensen}, F.~E., {et~al.} 2013, \apj,
  770, 103

\bibitem[{{HI4PI Collaboration} {et~al.}(2016){HI4PI Collaboration}, {Ben
  Bekhti}, {Fl{\"o}er}, {Keller}, {Kerp}, {Lenz}, {Winkel}, {Bailin},
  {Calabretta}, {Dedes}, {Ford}, {Gibson}, {Haud}, {Janowiecki}, {Kalberla},
  {Lockman}, {McClure-Griffiths}, {Murphy}, {Nakanishi}, {Pisano}, \&
  {Staveley-Smith}}]{HI4PI}
{HI4PI Collaboration}, {Ben Bekhti}, N., {Fl{\"o}er}, L., {et~al.} 2016, \aap,
  594, A116

\bibitem[{{Jiang} {et~al.}(2019){Jiang}, {Fabian}, {Dauser}, {Gallo},
  {Garc{\'\i}a}, {Kara}, {Parker}, {Tomsick}, {Walton}, \&
  {Reynolds}}]{JiangJ19}
{Jiang}, J., {Fabian}, A.~C., {Dauser}, T., {et~al.} 2019, \mnras, 489, 3436

\bibitem[{{Kang} {et~al.}(2021){Kang}, {Wang}, \& {Kang}}]{Kang21}
{Kang}, J.-L., {Wang}, J.-X., \& {Kang}, W.-Y. 2021, \mnras, 502, 80

\bibitem[{{Kara} {et~al.}(2017){Kara}, {Garc{\'{\i}}a}, {Lohfink}, {Fabian},
  {Reynolds}, {Tombesi}, \& {Wilkins}}]{Kara17}
{Kara}, E., {Garc{\'{\i}}a}, J.~A., {Lohfink}, A., {et~al.} 2017, \mnras, 468,
  3489

\bibitem[{{Keek} \& {Ballantyne}(2016)}]{Keek16}
{Keek}, L. \& {Ballantyne}, D.~R. 2016, \mnras, 456, 2722

\bibitem[{{Kollatschny}(2003)}]{Kollatschny03}
{Kollatschny}, W. 2003, \aap, 412, L61

\bibitem[{{Kubota} \& {Done}(2018)}]{Kubota18}
{Kubota}, A. \& {Done}, C. 2018, \mnras, 480, 1247

\bibitem[{{Liu} {et~al.}(2017){Liu}, {Feng}, \& {Bai}}]{Liu17}
{Liu}, H.~T., {Feng}, H.~C., \& {Bai}, J.~M. 2017, \mnras, 466, 3323

\bibitem[{{Liu} \& {Wang}(2010)}]{Liu10}
{Liu}, T. \& {Wang}, J.-X. 2010, \apj, 725, 2381

\bibitem[{{Lohfink} {et~al.}(2016){Lohfink}, {Reynolds}, {Pinto}, {Alston},
  {Boggs}, {Christensen}, {Craig}, {Fabian}, {Hailey}, {Harrison}, {Kara},
  {Matt}, {Parker}, {Stern}, {Walton}, \& {Zhang}}]{Lohfink16}
{Lohfink}, A.~M., {Reynolds}, C.~S., {Pinto}, C., {et~al.} 2016, \apj, 821, 11

\bibitem[{{Magdziarz} {et~al.}(1998){Magdziarz}, {Blaes}, {Zdziarski},
  {Johnson}, \& {Smith}}]{Magdziarz98}
{Magdziarz}, P., {Blaes}, O.~M., {Zdziarski}, A.~A., {Johnson}, W.~N., \&
  {Smith}, D.~A. 1998, \mnras, 301, 179

\bibitem[{{Mantovani} {et~al.}(2016){Mantovani}, {Nandra}, \&
  {Ponti}}]{Mantovani16}
{Mantovani}, G., {Nandra}, K., \& {Ponti}, G. 2016, \mnras, 458, 4198

\bibitem[{{Marinucci} {et~al.}(2019){Marinucci}, {Porquet}, {Tamborra},
  {Bianchi}, {Braito}, {Lobban}, {Marin}, {Matt}, {Middei}, {Nardini},
  {Reeves}, \& {Tortosa}}]{Marinucci19}
{Marinucci}, A., {Porquet}, D., {Tamborra}, F., {et~al.} 2019, \aap, 623, A12

\bibitem[{{Matt} {et~al.}(2014){Matt}, {Marinucci}, {Guainazzi}, {Brenneman},
  {Elvis}, {Lohfink}, {Ar{\`e}valo}, {Boggs}, {Cappi}, {Christensen}, {Craig},
  {Fabian}, {Fuerst}, {Hailey}, {Harrison}, {Parker}, {Reynolds}, {Stern},
  {Walton}, \& {Zhang}}]{Matt14}
{Matt}, G., {Marinucci}, A., {Guainazzi}, M., {et~al.} 2014, \mnras, 439, 3016

\bibitem[{{Matzeu} {et~al.}(2020){Matzeu}, {Nardini}, {Parker}, {Reeves},
  {Braito}, {Porquet}, {Middei}, {Kammoun}, {Lusso}, {Alston}, {Giustini},
  {Lobban}, {Joyce}, {Igo}, {Ebrero}, {Ballo}, {Santos-Lle{\'o}}, \&
  {Schartel}}]{Matzeu20}
{Matzeu}, G.~A., {Nardini}, E., {Parker}, M.~L., {et~al.} 2020, \mnras, 497,
  2352

\bibitem[{{Mehdipour} {et~al.}(2015){Mehdipour}, {Kaastra}, {Kriss}, {Cappi},
  {Petrucci}, {Steenbrugge}, {Arav}, {Behar}, {Bianchi}, {Boissay},
  {Branduardi-Raymont}, {Costantini}, {Ebrero}, {Di Gesu}, {Harrison}, {Kaspi},
  {De Marco}, {Matt}, {Paltani}, {Peterson}, {Ponti}, {Pozo Nu{\~n}ez}, {De
  Rosa}, {Ursini}, {de Vries}, {Walton}, \& {Whewell}}]{Mehdipour15}
{Mehdipour}, M., {Kaastra}, J.~S., {Kriss}, G.~A., {et~al.} 2015, \aap, 575,
  A22

\bibitem[{{Middei} {et~al.}(2019){Middei}, {Bianchi}, {Petrucci}, {Ursini},
  {Cappi}, {De Marco}, {De Rosa}, {Malzac}, {Marinucci}, {Matt}, {Ponti}, \&
  {Tortosa}}]{Middei19}
{Middei}, R., {Bianchi}, S., {Petrucci}, P.~O., {et~al.} 2019, \mnras, 483,
  4695

\bibitem[{{Middei} {et~al.}(2020){Middei}, {Petrucci}, {Bianchi}, {Ursini},
  {Cappi}, {Clavel}, {De Rosa}, {Marinucci}, {Matt}, \& {Tortosa}}]{Middei20}
{Middei}, R., {Petrucci}, P.~O., {Bianchi}, S., {et~al.} 2020, \aap, 640, A99

\bibitem[{{Murphy} \& {Yaqoob}(2009)}]{Murphy09}
{Murphy}, K.~D. \& {Yaqoob}, T. 2009, \mnras, 397, 1549

\bibitem[{{Nied{\'z}wiecki} {et~al.}(2019){Nied{\'z}wiecki}, {Szanecki}, \&
  {Zdziarski}}]{Niedzwiecki19}
{Nied{\'z}wiecki}, A., {Szanecki}, M., \& {Zdziarski}, A.~A. 2019, \mnras, 485,
  2942

\bibitem[{{Osterbrock} \& {Pogge}(1985)}]{Osterbrock85}
{Osterbrock}, D.~E. \& {Pogge}, R.~W. 1985, \apj, 297, 166

\bibitem[{{Panagiotou} \& {Walter}(2020)}]{Panagiotou20}
{Panagiotou}, C. \& {Walter}, R. 2020, \aap, 640, A31

\bibitem[{{Patrick} {et~al.}(2012){Patrick}, {Reeves}, {Porquet}, {Markowitz},
  {Braito}, \& {Lobban}}]{Patrick12}
{Patrick}, A.~R., {Reeves}, J.~N., {Porquet}, D., {et~al.} 2012, \mnras, 426,
  2522

\bibitem[{{Petrucci} {et~al.}(2020){Petrucci}, {Gronkiewicz}, {Rozanska},
  {Belmont}, {Bianchi}, {Czerny}, {Matt}, {Malzac}, {Middei}, {De Rosa},
  {Ursini}, \& {Cappi}}]{Petrucci20}
{Petrucci}, P.~O., {Gronkiewicz}, D., {Rozanska}, A., {et~al.} 2020, \aap, 634,
  A85

\bibitem[{{Petrucci} {et~al.}(2013){Petrucci}, {Paltani}, {Malzac}, {Kaastra},
  {Cappi}, {Ponti}, {De Marco}, {Kriss}, {Steenbrugge}, {Bianchi},
  {Branduardi-Raymont}, {Mehdipour}, {Costantini}, {Dadina}, \&
  {Lubi{\'n}ski}}]{Petrucci13}
{Petrucci}, P.-O., {Paltani}, S., {Malzac}, J., {et~al.} 2013, \aap, 549, A73

\bibitem[{{Petrucci} {et~al.}(2018){Petrucci}, {Ursini}, {De Rosa}, {Bianchi},
  {Cappi}, {Matt}, {Dadina}, \& {Malzac}}]{Petrucci18}
{Petrucci}, P.-O., {Ursini}, F., {De Rosa}, A., {et~al.} 2018, \aap, 611, A59

\bibitem[{{Piconcelli} {et~al.}(2005){Piconcelli}, {Jimenez-Bail{\'o}n},
  {Guainazzi}, {Schartel}, {Rodr{\'{\i}}guez-Pascual}, \&
  {Santos-Lle{\'o}}}]{Piconcelli05}
{Piconcelli}, E., {Jimenez-Bail{\'o}n}, E., {Guainazzi}, M., {et~al.} 2005,
  \aap, 432, 15

\bibitem[{{Planck Collaboration} {et~al.}(2020){Planck Collaboration},
  {Aghanim}, {Akrami}, {Ashdown}, {Aumont}, {Baccigalupi}, {Ballardini},
  {Banday}, {Barreiro}, {Bartolo}, {Basak}, {Battye}, {Benabed}, {Bernard},
  {Bersanelli}, {Bielewicz}, {Bock}, {Bond}, {Borrill}, {Bouchet}, {Boulanger},
  {Bucher}, {Burigana}, {Butler}, {Calabrese}, {Cardoso}, {Carron},
  {Challinor}, {Chiang}, {Chluba}, {Colombo}, {Combet}, {Contreras}, {Crill},
  {Cuttaia}, {de Bernardis}, {de Zotti}, {Delabrouille}, {Delouis}, {Di
  Valentino}, {Diego}, {Dor{\'e}}, {Douspis}, {Ducout}, {Dupac}, {Dusini},
  {Efstathiou}, {Elsner}, {En{\ss}lin}, {Eriksen}, {Fantaye}, {Farhang},
  {Fergusson}, {Fernandez-Cobos}, {Finelli}, {Forastieri}, {Frailis},
  {Fraisse}, {Franceschi}, {Frolov}, {Galeotta}, {Galli}, {Ganga},
  {G{\'e}nova-Santos}, {Gerbino}, {Ghosh}, {Gonz{\'a}lez-Nuevo}, {G{\'o}rski},
  {Gratton}, {Gruppuso}, {Gudmundsson}, {Hamann}, {Handley}, {Hansen},
  {Herranz}, {Hildebrandt}, {Hivon}, {Huang}, {Jaffe}, {Jones}, {Karakci},
  {Keih{\"a}nen}, {Keskitalo}, {Kiiveri}, {Kim}, {Kisner}, {Knox},
  {Krachmalnicoff}, {Kunz}, {Kurki-Suonio}, {Lagache}, {Lamarre}, {Lasenby},
  {Lattanzi}, {Lawrence}, {Le Jeune}, {Lemos}, {Lesgourgues}, {Levrier},
  {Lewis}, {Liguori}, {Lilje}, {Lilley}, {Lindholm}, {L{\'o}pez-Caniego},
  {Lubin}, {Ma}, {Mac{\'\i}as-P{\'e}rez}, {Maggio}, {Maino}, {Mandolesi},
  {Mangilli}, {Marcos-Caballero}, {Maris}, {Martin}, {Martinelli},
  {Mart{\'\i}nez-Gonz{\'a}lez}, {Matarrese}, {Mauri}, {McEwen}, {Meinhold},
  {Melchiorri}, {Mennella}, {Migliaccio}, {Millea}, {Mitra},
  {Miville-Desch{\^e}nes}, {Molinari}, {Montier}, {Morgante}, {Moss}, {Natoli},
  {N{\o}rgaard-Nielsen}, {Pagano}, {Paoletti}, {Partridge}, {Patanchon},
  {Peiris}, {Perrotta}, {Pettorino}, {Piacentini}, {Polastri}, {Polenta},
  {Puget}, {Rachen}, {Reinecke}, {Remazeilles}, {Renzi}, {Rocha}, {Rosset},
  {Roudier}, {Rubi{\~n}o-Mart{\'\i}n}, {Ruiz-Granados}, {Salvati}, {Sandri},
  {Savelainen}, {Scott}, {Shellard}, {Sirignano}, {Sirri}, {Spencer},
  {Sunyaev}, {Suur-Uski}, {Tauber}, {Tavagnacco}, {Tenti}, {Toffolatti},
  {Tomasi}, {Trombetti}, {Valenziano}, {Valiviita}, {Van Tent}, {Vibert},
  {Vielva}, {Villa}, {Vittorio}, {Wand elt}, {Wehus}, {White}, {White},
  {Zacchei}, \& {Zonca}}]{Planck20}
{Planck Collaboration}, {Aghanim}, N., {Akrami}, Y., {et~al.} 2020, \aap, 641,
  A6

\bibitem[{{Ponti} {et~al.}(2012){Ponti}, {Papadakis}, {Bianchi}, {Guainazzi},
  {Matt}, {Uttley}, \& {Bonilla}}]{Ponti12}
{Ponti}, G., {Papadakis}, I., {Bianchi}, S., {et~al.} 2012, \aap, 542, A83

\bibitem[{{Porquet} {et~al.}(2019){Porquet}, {Done}, {Reeves}, {Grosso},
  {Marinucci}, {Matt}, {Lobban}, {Nardini}, {Braito}, {Marin}, {Kubota},
  {Ricci}, {Koss}, {Stern}, {Ballantyne}, \& {Farrah}}]{Porquet19}
{Porquet}, D., {Done}, C., {Reeves}, J.~N., {et~al.} 2019, \aap, 623, A11

\bibitem[{{Porquet} {et~al.}(2018){Porquet}, {Reeves}, {Matt}, {Marinucci},
  {Nardini}, {Braito}, {Lobban}, {Ballantyne}, {Boggs}, {Christensen},
  {Dauser}, {Farrah}, {Garcia}, {Hailey}, {Harrison}, {Stern}, {Tortosa},
  {Ursini}, \& {Zhang}}]{Porquet18}
{Porquet}, D., {Reeves}, J.~N., {Matt}, G., {et~al.} 2018, \aap, 609, A42

\bibitem[{{Porquet} {et~al.}(2004){Porquet}, {Reeves}, {O'Brien}, \&
  {Brinkmann}}]{P04a}
{Porquet}, D., {Reeves}, J.~N., {O'Brien}, P., \& {Brinkmann}, W. 2004, \aap,
  422, 85

\bibitem[{{Poutanen} \& {Svensson}(1996)}]{Poutanen96}
{Poutanen}, J. \& {Svensson}, R. 1996, \apj, 470, 249

\bibitem[{{Reeves} {et~al.}(2021{\natexlab{a}}){Reeves}, {Braito}, {Porquet},
  {Lobban}, {Matzeu}, \& {Nardini}}]{Reeves21a}
{Reeves}, J.~N., {Braito}, V., {Porquet}, D., {et~al.} 2021{\natexlab{a}},
  \mnras, 500, 1974

\bibitem[{{Reeves} {et~al.}(2021{\natexlab{b}}){Reeves}, {Porquet}, {Braito},
  {Grosso}, \& {Lobban}}]{Reeves21b}
{Reeves}, J.~N., {Porquet}, D., {Braito}, V., {Grosso}, N., \& {Lobban}, A.
  2021{\natexlab{b}}, \aap, 649, L3

\bibitem[{{Ricci} {et~al.}(2014){Ricci}, {Ueda}, {Ichikawa}, {Paltani},
  {Boissay}, {Gandhi}, {Stalevski}, \& {Awaki}}]{Ricci14}
{Ricci}, C., {Ueda}, Y., {Ichikawa}, K., {et~al.} 2014, \aap, 567, A142

\bibitem[{{Schnittman} \& {Krolik}(2010)}]{Schnittman10}
{Schnittman}, J.~D. \& {Krolik}, J.~H. 2010, \apj, 712, 908

\bibitem[{{Shu} {et~al.}(2010){Shu}, {Yaqoob}, \& {Wang}}]{Shu10}
{Shu}, X.~W., {Yaqoob}, T., \& {Wang}, J.~X. 2010, \apjs, 187, 581

\bibitem[{{Steiner} {et~al.}(2017){Steiner}, {Garc{\'{\i}}a}, {Eikmann},
  {McClintock}, {Brenneman}, {Dauser}, \& {Fabian}}]{Steiner17}
{Steiner}, J.~F., {Garc{\'{\i}}a}, J.~A., {Eikmann}, W., {et~al.} 2017, \apj,
  836, 119

\bibitem[{{Str{\"u}der} {et~al.}(2001){Str{\"u}der}, {Briel}, {Dennerl},
  {Hartmann}, {Kendziorra}, {Meidinger}, {Pfeffermann}, {Reppin}, {Aschenbach},
  {Bornemann}, {Br{\"a}uninger}, {Burkert}, {Elender}, {Freyberg}, {Haberl},
  {Hartner}, {Heuschmann}, {Hippmann}, {Kastelic}, {Kemmer}, {Kettenring},
  {Kink}, {Krause}, {M{\"u}ller}, {Oppitz}, {Pietsch}, {Popp}, {Predehl},
  {Read}, {Stephan}, {St{\"o}tter}, {Tr{\"u}mper}, {Holl}, {Kemmer}, {Soltau},
  {St{\"o}tter}, {Weber}, {Weichert}, {von Zanthier}, {Carathanassis}, {Lutz},
  {Richter}, {Solc}, {B{\"o}ttcher}, {Kuster}, {Staubert}, {Abbey}, {Holland},
  {Turner}, {Balasini}, {Bignami}, {La Palombara}, {Villa}, {Buttler},
  {Gianini}, {Lain{\'e}}, {Lumb}, \& {Dhez}}]{Struder01}
{Str{\"u}der}, L., {Briel}, U., {Dennerl}, K., {et~al.} 2001, \aap, 365, L18

\bibitem[{{Tamborra} {et~al.}(2018){Tamborra}, {Matt}, {Bianchi}, \&
  {Dov{\v{c}}iak}}]{Tamborra18}
{Tamborra}, F., {Matt}, G., {Bianchi}, S., \& {Dov{\v{c}}iak}, M. 2018, \aap,
  619, A105

\bibitem[{{Titarchuk}(1994)}]{Titarchuk94}
{Titarchuk}, L. 1994, \apj, 434, 570

\bibitem[{{Tortosa} {et~al.}(2017){Tortosa}, {Marinucci}, {Matt}, {Bianchi},
  {La Franca}, {Ballantyne}, {Boorman}, {Fabian}, {Farrah}, {Fuerst}, {Gandhi},
  {Harrison}, {Koss}, {Ricci}, {Stern}, {Ursini}, \& {Walton}}]{Tortosa17}
{Tortosa}, A., {Marinucci}, A., {Matt}, G., {et~al.} 2017, \mnras, 466, 4193

\bibitem[{{Ursini} {et~al.}(2020){Ursini}, {Petrucci}, {Bianchi}, {Matt},
  {Middei}, {Marcel}, {Ferreira}, {Cappi}, {De Marco}, {De Rosa}, {Malzac},
  {Marinucci}, {Ponti}, \& {Tortosa}}]{Ursini20}
{Ursini}, F., {Petrucci}, P.~O., {Bianchi}, S., {et~al.} 2020, \aap, 634, A92

\bibitem[{{Ursini} {et~al.}(2016){Ursini}, {Petrucci}, {Matt}, {Bianchi},
  {Cappi}, {De Marco}, {De Rosa}, {Malzac}, {Marinucci}, {Ponti}, \&
  {Tortosa}}]{Ursini16}
{Ursini}, F., {Petrucci}, P.-O., {Matt}, G., {et~al.} 2016, \mnras, 463, 382

\bibitem[{{Verner} {et~al.}(1996){Verner}, {Ferland}, {Korista}, \&
  {Yakovlev}}]{Verner96}
{Verner}, D.~A., {Ferland}, G.~J., {Korista}, K.~T., \& {Yakovlev}, D.~G. 1996,
  \apj, 465, 487

\bibitem[{{V{\'e}ron-Cetty} {et~al.}(2007){V{\'e}ron-Cetty}, {V{\'e}ron},
  {Joly}, \& {Kollatschny}}]{Veron-cetty07}
{V{\'e}ron-Cetty}, M.~P., {V{\'e}ron}, P., {Joly}, M., \& {Kollatschny}, W.
  2007, \aap, 475, 487

\bibitem[{{Vincentelli} {et~al.}(2021){Vincentelli}, {McHardy}, {Cackett},
  {Barth}, {Horne}, {Goad}, {Korista}, {Gelbord}, {Brandt}, {Edelson},
  {Miller}, {Pahari}, {Peterson}, {Schmidt}, {Baldi}, {Breedt}, {Hern{\'a}ndez
  Santisteban}, {Romero-Colmenero}, {Ward}, \& {Williams}}]{Vincentelli21}
{Vincentelli}, F.~M., {McHardy}, I., {Cackett}, E.~M., {et~al.} 2021, \mnras,
  504, 4337

\bibitem[{{Waddell} \& {Gallo}(2020)}]{Waddel20}
{Waddell}, S.~G.~H. \& {Gallo}, L.~C. 2020, \mnras, 498, 5207

\bibitem[{{Walton} {et~al.}(2020){Walton}, {Alston}, {Kosec}, {Fabian},
  {Gallo}, {Garcia}, {Miller}, {Nardini}, {Reynolds}, {Ricci}, {Stern},
  {Dauser}, {Harrison}, \& {Reynolds}}]{Walton20}
{Walton}, D.~J., {Alston}, W.~N., {Kosec}, P., {et~al.} 2020, \mnras, 499, 1480

\bibitem[{{Walton} {et~al.}(2013){Walton}, {Nardini}, {Fabian}, {Gallo}, \&
  {Reis}}]{Walton13}
{Walton}, D.~J., {Nardini}, E., {Fabian}, A.~C., {Gallo}, L.~C., \& {Reis},
  R.~C. 2013, \mnras, 428, 2901

\bibitem[{{Weisskopf} {et~al.}(2016){Weisskopf}, {Ramsey}, {O'Dell}, {Tennant},
  {Elsner}, {Soffitta}, {Bellazzini}, {Costa}, {Kolodziejczak}, {Kaspi},
  {Muleri}, {Marshall}, {Matt}, \& {Romani}}]{Weisskopf16}
{Weisskopf}, M.~C., {Ramsey}, B., {O'Dell}, S., {et~al.} 2016, in \procspie,
  Vol. 9905, Space Telescopes and Instrumentation 2016: Ultraviolet to Gamma
  Ray, 990517

\bibitem[{{Williams} {et~al.}(2018){Williams}, {Gliozzi}, \&
  {Rudzinsky}}]{Williams18}
{Williams}, J.~K., {Gliozzi}, M., \& {Rudzinsky}, R.~V. 2018, \mnras, 480, 96

\bibitem[{{Wilms} {et~al.}(2000){Wilms}, {Allen}, \& {McCray}}]{Wilms00}
{Wilms}, J., {Allen}, A., \& {McCray}, R. 2000, \apj, 542, 914

\bibitem[{{XRISM Science Team}(2020)}]{XRISM20}
{XRISM Science Team}. 2020, arXiv e-prints, arXiv:2003.04962

\bibitem[{{Xu} {et~al.}(2021){Xu}, {Garc{\'\i}a}, {Walton}, {Connors},
  {Madsen}, \& {Harrison}}]{XuY21}
{Xu}, Y., {Garc{\'\i}a}, J.~A., {Walton}, D.~J., {et~al.} 2021, \apj, 913, 13

\bibitem[{{Zdziarski} {et~al.}(1996){Zdziarski}, {Johnson}, \&
  {Magdziarz}}]{Zdziarski96}
{Zdziarski}, A.~A., {Johnson}, W.~N., \& {Magdziarz}, P. 1996, \mnras, 283, 193

\bibitem[{{Zdziarski} {et~al.}(2020){Zdziarski}, {Szanecki}, {Poutanen},
  {Gierli{\'n}ski}, \& {Biernacki}}]{Zdziarski20}
{Zdziarski}, A.~A., {Szanecki}, M., {Poutanen}, J., {Gierli{\'n}ski}, M., \&
  {Biernacki}, P. 2020, \mnras, 492, 5234

\bibitem[{{Zhang} {et~al.}(2018){Zhang}, {Wang}, \& {Zhu}}]{ZhangJ18}
{Zhang}, J.-X., {Wang}, J.-X., \& {Zhu}, F.-F. 2018, \apj, 863, 71

\bibitem[{{Zhang} {et~al.}(2016){Zhang}, {Feroci}, {Santangelo}, {Dong},
  {Feng}, {Lu}, {Nandra}, {Wang}, {Zhang}, {Bozzo}, {Brandt}, {De Rosa}, {Gou},
  {Hernanz}, {van der Klis}, {Li}, {Liu}, {Orleanski}, {Pareschi}, {Pohl},
  {Poutanen}, {Qu}, {Schanne}, {Stella}, {Uttley}, {Watts}, {Xu}, {Yu}, {in 't
  Zand}, {Zane}, {Alvarez}, {Amati}, {Baldini}, {Bambi}, {Basso},
  {Bhattacharyya S.}, {}, {Belloni}, {Bellutti}, {Bianchi}, {Brez}, {Bursa},
  {Burwitz}, {Budtz-J{\o}rgensen}, {Caiazzo}, {Campana}, {Cao}, {Casella},
  {Chen}, {Chen}, {Chen}, {Chen}, {Chen}, {Chen}, {Civitani}, {Coti Zelati},
  {Cui}, {Cui}, {Dai}, {Del Monte}, {de Martino}, {Di Cosimo}, {Diebold},
  {Dovciak}, {Donnarumma}, {Doroshenko}, {Esposito}, {Evangelista}, {Favre},
  {Friedrich}, {Fuschino}, {Galvez}, {Gao}, {Ge}, {Gevin}, {Goetz}, {Han},
  {Heyl}, {Horak}, {Hu}, {Huang}, {Huang}, {Hudec}, {Huppenkothen}, {Israel},
  {Ingram}, {Karas}, {Karelin}, {Jenke}, {Ji}, {Korpela}, {Kunneriath},
  {Labanti}, {Li}, {Li}, {Li}, {Liang}, {Limousin}, {Lin}, {Ling}, {Liu},
  {Liu}, {Liu}, {Lu}, {Lund}, {Lai}, {Luo}, {Luo}, {Ma}, {Mahmoodifar},
  {Marisaldi}, {Martindale}, {Meidinger}, {Men}, {Michalska}, {Mignani},
  {Minuti}, {Motta}, {Muleri}, {Neilsen}, {Orlandini}, {Pan}, {Patruno},
  {Perinati}, {Picciotto}, {Piemonte}, {Pinchera}, {Rachevski A.}, {Rapisarda},
  {Rea}, {Rossi}, {Rubini}, {Sala}, {Shu}, {Sgro}, {Shen}, {Soffitta}, {Song},
  {Spandre}, {Stratta}, {Strohmayer}, {Sun}, {Svoboda}, {Tagliaferri},
  {Tenzer}, {Hong}, {Taverna}, {Torok}, {Turolla}, {Vacchi}, {Wang}, {Walton},
  {Wang}, {Wang}, {Wang}, {Wang}, {Weng}, {Wilms}, {Winter}, {Wu}, {Wu},
  {Xiong}, {Xu}, {Xue}, {Yan}, {Yang}, {Yang}, {Yang}, {Yuan}, {Yuan}, {Yuan},
  {Zampa}, {Zampa}, {Zdziarski}, {Zhang}, {Zhang}, {Zhang}, {Zhang}, {Zhang},
  {Zhang}, {Zheng}, {Zhou}, \& {Zhou X.~L.}}]{ZhangS16}
{Zhang}, S.~N., {Feroci}, M., {Santangelo}, A., {et~al.} 2016, in Society of
  Photo-Optical Instrumentation Engineers (SPIE) Conference Series, Vol. 9905,
  Space Telescopes and Instrumentation 2016: Ultraviolet to Gamma Ray, ed.
  J.-W.~A. {den Herder}, T.~{Takahashi}, \& M.~{Bautz}, 99051Q

\bibitem[{{Zhou} \& {Zhang}(2010)}]{Zhou10b}
{Zhou}, X.-L. \& {Zhang}, S.-N. 2010, \apjl, 713, L11

\bibitem[{{{\.Z}ycki} {et~al.}(1999){{\.Z}ycki}, {Done}, \& {Smith}}]{Zycki99}
{{\.Z}ycki}, P.~T., {Done}, C., \& {Smith}, D.~A. 1999, \mnras, 309, 561

\end{thebibliography}

\appendix

\section{Results of relativistic reflection modelling over the full X-ray broad-band range}\label{sec:appendix}

In this appendix, we report the spectral analysis of the two simultaneous {\sl XMM-Newton} and {\sl NuSTAR} observations, 
considering only relativistic reflection models.  

\begin{figure}[t!]
\begin{tabular}{c}
\includegraphics[width=0.9\columnwidth,angle=0]{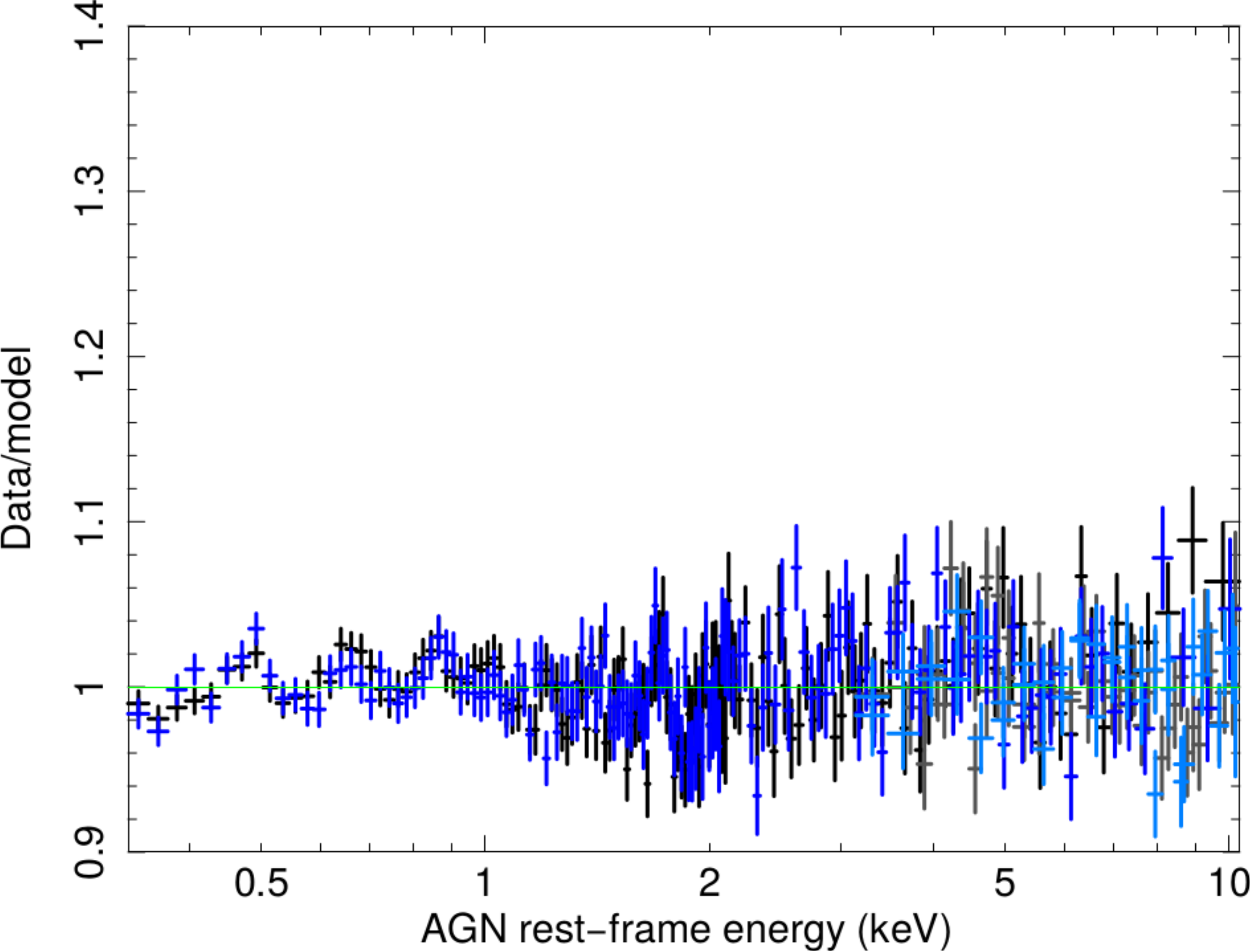}\\
\end{tabular}
\caption{Data/model ratio of the two simultaneous 2019 and 2020 {\sl XMM-Newton}/pn and {\sl NuSTAR} spectra over the 0.3--10\,keV energy range fit with the {\sc relxill} model. 
The Y-axis range is identical to that of Fig.~\ref{fig:refl} for a direct comparison. 
Black: 2019 {\sl XMM-Newton}/pn, dark grey: 2019 {\sl NuSTAR}, 
 blue: 2020 {\sl XMM-Newton}/pn, light blue: 2020 {\sl NuSTAR}. 
See Table~\ref{tab:refl} for the corresponding inferred parameter values.}  
\label{fig:reflbelow10keV}
\end{figure}

The fit using the {\sc relxill} model between 0.3 and 10\,keV provides a good statistical fit (Table~\ref{tab:refl}, Fig.~\ref{fig:reflbelow10keV}). 
However, it fails to reproduce the hard X-ray shape when data above 10\,keV are included in the fit 
(Fig.~\ref{fig:refl}, bottom panel), 
 with very significant positive hard X-ray residuals for both observations above $\sim$30\,keV 
 ($\chi^{2}$/d.o.f.=1954.8/1617, $\chi^{2}_{\rm red}$ = 1.21).  
 Since the fit is driven by the smooth soft X-ray emission, high values for 
 the disc emissivity index, the photon power-law index and the reflection strength 
 are required to reproduce it (Table~\ref{tab:refl}, third column), 
 at odds with the low values required to explain the data above 3\,keV (Table~\ref{tab:above3keV}). 
 
We then investigate the scenario whereby the incident continuum of the relativistic 
component can be different from the direct observed one, as was proposed by \cite{Fuerst15}
(though see also \citealt{Steiner17}) 
to explain the X-ray spectrum of the X-ray binary \object{GX 339-4} in low-luminosity, hard states. 
However, this requires a very significant divergence between the incident and observed spectral indices 
with $\Delta\Gamma$=0.7 (at 22$\sigma$ confidence level) 
 and $\Delta\Gamma$=0.9 (at 14$\sigma$ confidence level) for the 2019 and 2020 observations, respectively. 
 This means that this scenario appears highly unlikely to explain these observations of Mrk\,110. 
 
Alternatively, we investigate whether the broad-band X-ray spectra can be explained by a 
lamppost geometry for the corona, 
different disc properties (disc ionization gradient, density)
and/or a different continuum shape (Comptonisation). 
Even relativistic reflection from two different medium properties, allowing for different ionisation parameters, have been investigated. 
 The modelling is performed thanks to the various available relxill model flavors. 
 However, none of them is able to simultaneously account for the soft X-ray excess and the hard X-ray emission shape 
 in both the 2019 and 2020 {\sl XMM-Newton} and {\sl NuSTAR} observations.
Similar results are found using the relativistic modelling package {\sc reflkerr} \citep{Niedzwiecki19} 
assuming either a coronal geometry ({\sc reflkerr}) or a lamppost geometry ({\sc reflkerr\_lp}).
We note that even upon using an unblurred reflection component to take into account the very weak molecular torus contribution 
-- using {\sc xillver}, {\sc mytorus} \citep{Murphy09}, or {\sc borus12} \citep{Balokovic18,Balokovic19} --
does not help to remove these huge hard X-ray residuals.  \\

\begin{table}[t!]
\caption{Simultaneous fits of the two {\sl XMM-Newton} 
        and {\sl NuSTAR} spectra with a relativistic reflection 
	model with a coronal geometry for the accretion disc emissivity ({\sc relxill}). 
	The column density along the line-of-sight has been fixed at the Galactic value (1.27$\times$10$^{20}$\,cm$^{-2}$),
	and the accretion disc inclination angle at 9.9 degrees. 
	(a) The breaking radius $R_{\rm br}$ of the emissivity shape is expressed in ISCO units.
	(f) means that the cut-off value is fixed at that found from the fit above 3\,keV (see Table~\ref{tab:above3keV}).
}           
\centering                          
\begin{tabular}{lcc}        
\hline\hline                 
Parameters & \multicolumn{1}{c}{0.3--10\,keV} & \multicolumn{1}{c}{Full range} \\    
\hline                       
$a$         & 0.981$^{+0.008}_{-0.005}$ &  $\geq$0.996 \\
$A_{\rm Fe}$      & $\leq$0.53  &  3.4$^{+0.2}_{-0.1}$\\
\hline                                  
                          \multicolumn{3}{c}{2019 Nov} \\
\hline                                  
$q_{1}$ &  7.7$^{+0.5}_{-0.7}$   & 8.2$^{+0.1}_{-0.2}$  \\
$q_{2}$    & 2.0$\pm$0.2 & 3.4$\pm$0.1  \\
$R_{\rm br}$ ($R_{\rm isco}$)$^{\rm (a)}$ & 3.8$^{+1.1}_{-0.4}$  & 2.9$\pm$0.1    \\
$\Gamma$  & 2.28$\pm$0.01 & 1.99$^{+0.01}_{-0.02}$\\
$E_{\rm cut}$ (keV) &   113 (f)  & $\geq$941 \\
log\,$\xi$ (erg\,cm\,s$^{-1}$) & $\leq$0.04 & 3.0$\pm$0.1\\
$\mathcal{R}$       &  6.8$^{+0.5}_{-0.4}$  & 4.9$^{+0.3}_{-0.7}$\\
$norm_{\rm relxill}$ & 1.1$\pm0.1\times$10$^{-4}$ & 9.0$\pm$0.1$\times$10$^{-5}$ \\
\hline                                  
                 \multicolumn{3}{c}{2020 April} \\
\hline                                  
$q_{1}$  &  7.7$^{+0.6}_{-0.7}$   & 8.1$^{+0.1}_{-0.2}$     \\
$q_{2}$    & 2.2$\pm$0.2 & 3.4$\pm$0.1   \\
$R_{\rm br}$ ($R_{\rm isco}$)$^{\rm (a)}$ & 3.4$^{+0.9}_{-0.5}$  & 2.8$^{+0.1}_{-0.2}$   \\
$\Gamma$  & 2.16$^{+0.01}_{-0.02}$ &  1.90$\pm$0.01\\
$E_{\rm cut}$ (keV) &   126 (f) & $\geq$976\\
log\,$\xi$ (erg\,cm\,s$^{-1}$) & $\leq$0.09  & 3.0$\pm$0.1\\
$\mathcal{R}$ &  4.9$^{+0.4}_{-0.5}$  &  2.9$\pm$0.1\\
$norm_{\rm relxill}$  & 8.8$\pm$0.1$\times$10$^{-5}$ & 10.1$\pm$0.1$\times$10$^{-5}$  \\
 \hline
 \hline
$\chi^{2}$/d.o.f.  & 1181.9/1099 &  1954.8/1617\\
$\chi^{2}_{\rm red}$ &  1.08 &  1.21 \\
\hline    
\hline                  
\end{tabular}
\label{tab:refl}
\end{table}

\begin{figure}[t!]
\begin{tabular}{c}
\includegraphics[width=0.9\columnwidth,angle=0]{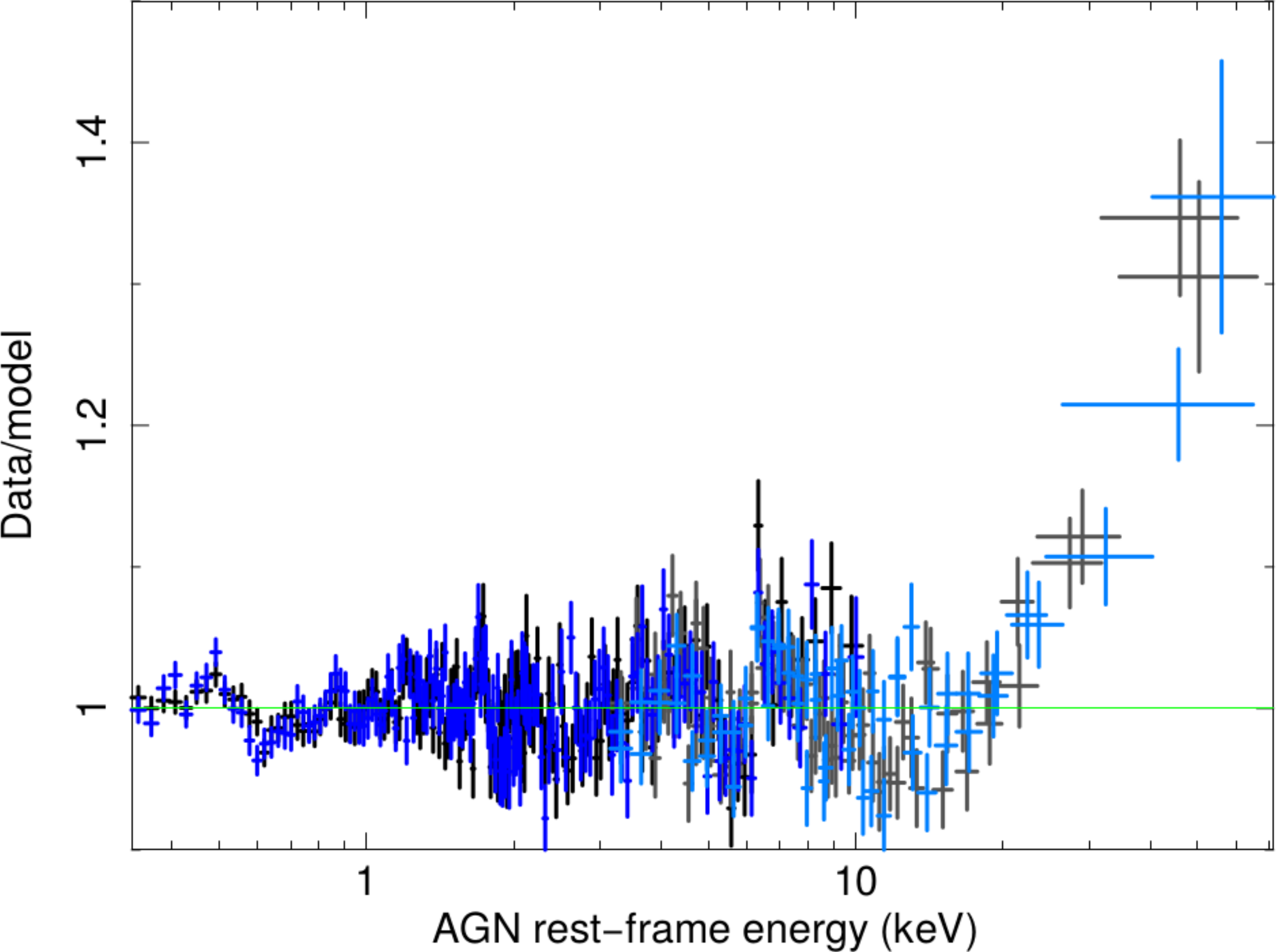}\\
\includegraphics[width=0.9\columnwidth,angle=0]{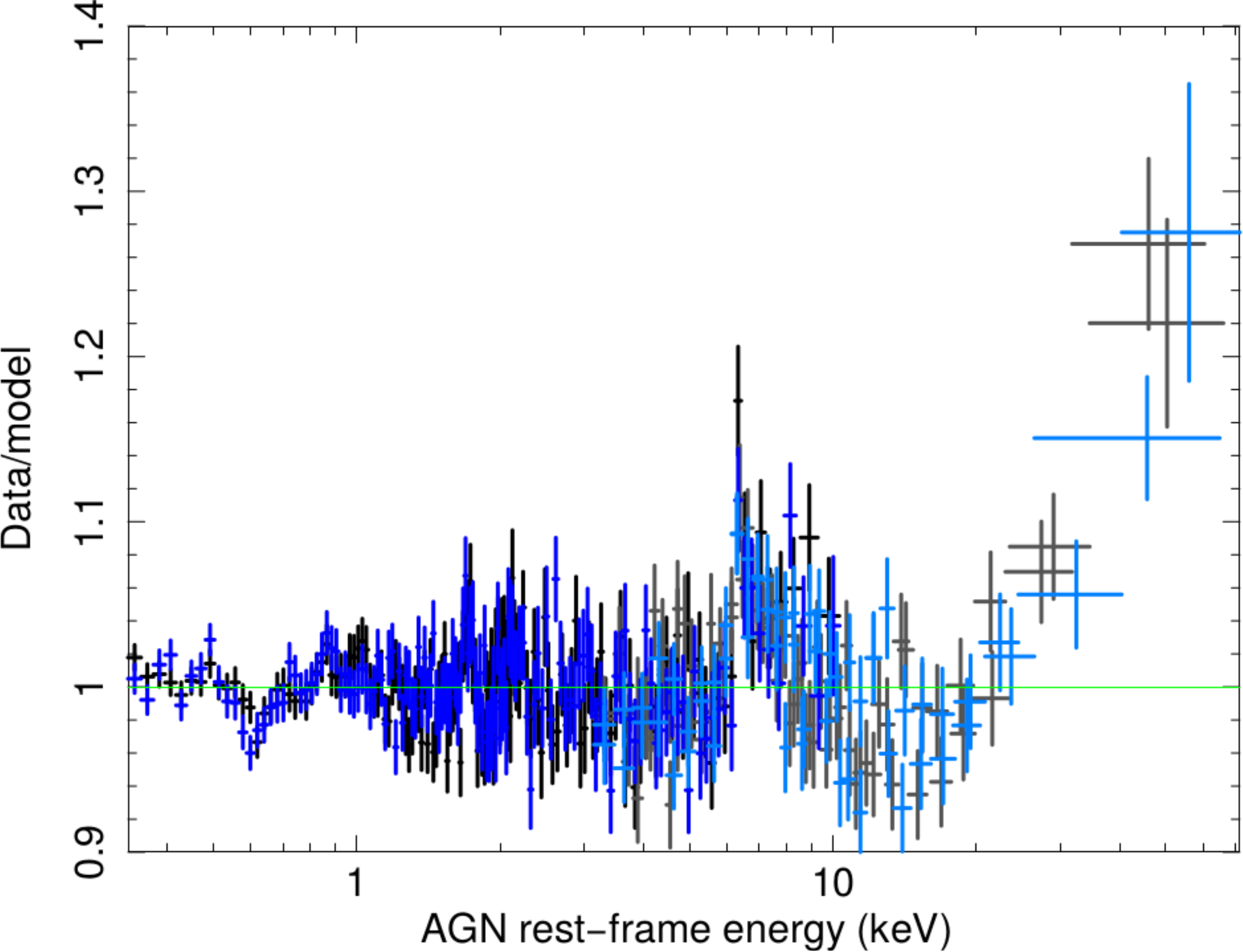}\\
\includegraphics[width=0.9\columnwidth,angle=0]{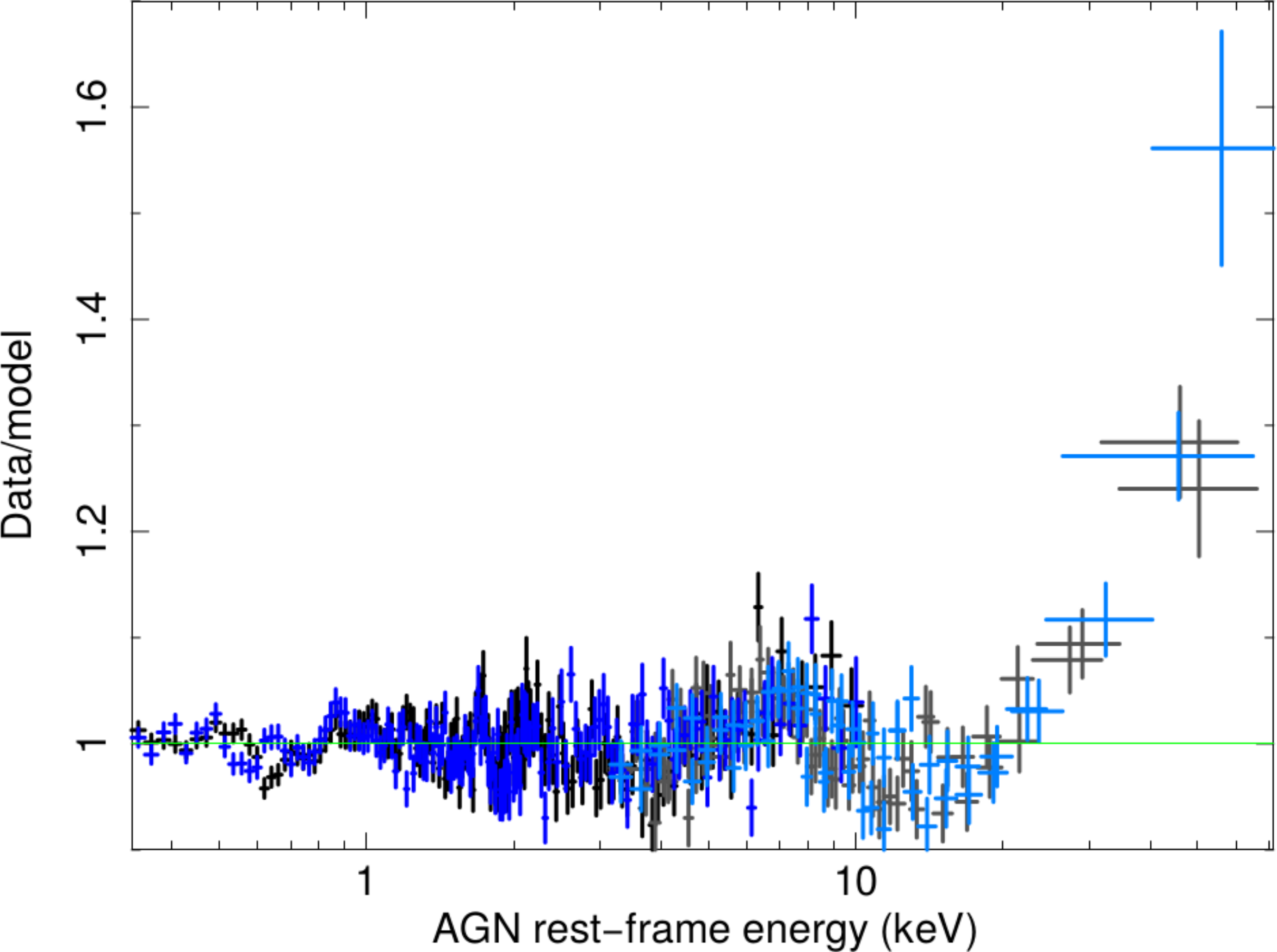}\\
\end{tabular}
\caption{Data/model ratio of the two simultaneous 2019 and 2020 {\sl XMM-Newton}/pn and {\sl NuSTAR} spectra fit with relativistic models. 
Black: 2019 {\sl XMM-Newton}/pn, dark grey: 2019 {\sl NuSTAR}, 
 blue: 2020 {\sl XMM-Newton}/pn, light blue: 2020 {\sl NuSTAR}. 
{\it Top panel}: {\sc relxilld} model. {\it Middle panel}: {\sc relxilllpion} model. {\it Bottom panel}: {\sc reflkerrd} model. 
See text for explanations, and Table~\ref{tab:otherrefl} for their corresponding inferred parameter values.}  
\label{fig:otherrefl}
\end{figure}

Some examples of the above models are described below for illustration purposes (Table~\ref{tab:otherrefl}, Fig.~\ref{fig:otherrefl}):
\begin{itemize}
 \item Higher disc density: 
As demonstrated by \cite{Garcia16} (see also \citealt{JiangJ19}) high-density accretion disc models can lead to a larger soft excess compared to those in lower density discs. Therefore, we investigate whether the strong soft X-ray excess found in Mrk\,110 can be explained by a higher accretion-disc density compared to the {\sc relxill} model, which assumes a fixed constant disc density of 10$^{15}$\,cm$^{-3}$. We use the {\sc relxilld} model, which allows the density to be free to vary between 10$^{15}$\,cm$^{-3}$ and 10$^{19}$\,cm$^{-3}$. For this model the cut-off energy value is fixed at 300\,keV and cannot be allowed to vary.
\item Lamppost geometry: 
The {\sc relxill} model assumes a coronal geometry for the disc emissivity. Here, we test the alternative disc emissivity, that is to say the lamppost geometry using {\sc relxilllpion}, with the X-ray point source being located at height $h$ above the black hole and outflowing at a velocity $\beta$=$v$/$c$. 
Moreover, this model allows for a disc ionization gradient along the radial distance with a power-law shape, $R^{\rm -index}$. 
\item The {\sc reflkerr} code: 
We also use the alternative relativistic reflection model \citep{Niedzwiecki19}. 
We apply the {\sc reflkerrd} model where the incident continuum is due to Comptonisation and a corona 
disc geometry is assumed. The disc density can vary between 10$^{15}$\,cm$^{-3}$ and 10$^{19}$\,cm$^{-3}$.
\end{itemize}

None of the above models provides a satisfactory account of both the soft and hard X-ray spectra of Mrk\,110. 
Indeed large hard X-ray excess are found (as per Fig.\ref{fig:otherrefl}) for all the cases considered and the models cannot simultaneously fit the soft excess, 
while at the same time correctly account for the hard X-ray spectral shape. 
Thus, it appears most likely that the broad band X-ray spectra of Mrk\,110 
require the presence of both soft and hard Comptonisation to explain the spectral curvature, 
as is discussed in the main text.

\begin{table}[t!]
\caption{Simultaneous fits of the two {\sl XMM-Newton} 
        and {\sl NuSTAR} spectra with three different relativistic reflection 
	models. The column density along the line-of-sight is fixed at the Galactic one, that is, 1.27$\times$10$^{20}$\,cm$^{-2}$, and the inclination angle at 9.9 degrees. 
	(a) Velocity of primary source. 
	(b) Index of the ionisation gradient with a power-law shape.
	(c) The breaking radius $R_{\rm br}$ and the height of the primary source are both expressed in ISCO units, except for the {\sc reflkerr} model for which $R_{\rm br}$ is expressed in gravitational radius units.
	The data-to-model ratio for each model is reported in Fig.~\ref{fig:otherrefl}.
}           
\centering                          
\begin{tabular}{@{}lc@{}c@{}c@{}}        
\hline\hline                 
Parameters & \multicolumn{1}{c}{{\sc relxilld}} & \multicolumn{1}{c}{{\sc relxilllpion}}  & \multicolumn{1}{c}{{\sc reflkerrd}} \\    
\hline                       
$a$         & 0.993$\pm$0.001 &  0.997$\pm$0.001  & $\geq$0.991\\
$A_{\rm Fe}$      & 4.5$\pm$0.1 & 2.1$^{+0.3}_{-0.2}$ & 2.9$\pm$0.2\\
log\,N$_{\rm disc}$ (cm$^{-3}$) & 17.0$\pm$0.1 & $-$ & 17.1$\pm$0.1 \\
$\beta$ ~~(${\rm v/c}$)$^{\rm (a)}$ & $-$ & 0.07$\pm$0.01 & $-$ \\
$\xi-index$$^{\rm (b)}$    & $-$   & $\leq$0.02   & $-$   \\
\hline                                  
                          \multicolumn{4}{c}{2019 Nov} \\
\hline                                  
\vspace{0.5mm}
$q_{1}$     & 8.3$\pm$0.1  & $-$ & 4.8$\pm$0.1 \\
$q_{2}$    &  3.4$\pm$0.1  & $-$ & 2.3$\pm$0.1 \\   
$R_{\rm br}$$^{\rm (c)}$ &  2.6$\pm$0.1 & $-$ &  13.4$^{+3.7}_{-1.0}$  \\
$h$$^{\rm (c)}$  & $-$ & $\leq$1.2 & $-$\\
$\Gamma$  & 1.92$\pm$0.01 & 2.01$\pm$0.01 & $-$\\
$kT_{\rm hot}$ (keV) & $-$ & $-$ & 88$\pm$2  \\ 
$y$ & $-$ & $-$ & 0.44$\pm$0.01 \\
$E_{\rm cut}$ (keV) & 300 (f)   & $\geq$872  &  $-$ \\
log\,$\xi$ (erg\,cm\,s$^{-1}$) & 3.0$\pm$0.1 & 2.7$\pm$0.1 & 2.6$\pm$0.1\\
$\mathcal{R}$       &  3.8$\pm$0.1  & $-$ & 1.2$\pm$0.1 \\
$norm_{\rm rel}$ & 9.0$\pm0.1\times$10$^{-5}$ & 0.28$\pm$0.1 &  6.6$\pm$0.1$\times$10$^{-3}$ \\
\hline                                  
                 \multicolumn{4}{c}{2020 April} \\
\hline                                  
$q_{1}$  &  8.0$\pm$0.1   & $-$ & 4.5$\pm$0.1     \\
$q_{2}$    &  3.3$\pm$0.1 & $-$ &  2.1$\pm$0.1   \\
$R_{\rm br}$${\rm (c)}$ & 2.7$\pm$0.1 & $-$ & 10.3$^{+1.2}_{-1.8}$    \\
$h$$^{\rm (c)}$  & $-$ & $\leq$1.1 & $-$\\
$\Gamma$  & 1.84$\pm$0.01 &  1.88$\pm$0.01  & $-$\\
$kT_{\rm hot}$ (keV) & $-$ & $-$ & 76$\pm$2 \\ 
$y$ & $-$ & $-$ &  0.37$\pm$0.01\\ 
$E_{\rm cut}$ (keV) & 300 (f) &   451$^{+102}_{-71}$ & $-$\\
log\,$\xi$ (erg\,cm\,s$^{-1}$) & 3.0$\pm$0.1 &2.9$\pm$0.1 & 0.3$\pm$0.1\\
$\mathcal{R}$ &  2.6$\pm$0.1 & $-$ &  2.2$\pm$0.1\\
$norm_{\rm rel}$  & 9.3$\pm$0.1$\times$10$^{-5}$ & 0.42$^{+0.13}_{-0.03}$ & 7.3$\pm$0.1$\times$10$^{-3}$  \\
 \hline
 \hline
$\chi^{2}$/d.o.f.  & 2001.7/1618 &  2045.9/1621 & 2065.4/1616\\
$\chi^{2}_{\rm red}$ & 1.24  & 1.26  & 1.28\\
\hline    
\hline                  
\end{tabular}
\label{tab:otherrefl}
\end{table}

\end{document}